\documentclass[10pt,twocolumn,english,aps,preprint,prb,floatfix,superscriptaddress,twocolumn,showpacs,amsmath,amssymb,reprint]{revtex4-1}
\usepackage[T1]{fontenc}
\usepackage[latin9]{inputenc}
\usepackage{color}
\usepackage{graphicx}
\usepackage{esint}
\usepackage[caption=false]{subfig}
\usepackage[colorlinks=true,citecolor=blue,urlcolor=blue]{hyperref}
\providecommand{\tabularnewline}{\\}

\begin{document}

\title{Topological strings linking with quasi-particle exchange in superconducting Dirac semimetals}

\author{Pedro L. e S. Lopes}
\email{pls.lopes@usherbrooke.ca}
\affiliation{D\'{e}partement de physique and Institut Quantique, Universit\'e de Sherbrooke, Sherbrooke, Qu\'ebec, Canada J1K 2R1}

\author{Jeffrey C. Y. Teo}
\email{jteo@virginia.edu}
\affiliation{Department of Physics, University of Virginia, VA22904, USA}

\author{Shinsei Ryu}
\email{ryuu@uchicago.edu}
\affiliation{James Franck Institute and Kadanoff Center for Theoretical Physics,
University of Chicago, Illinois 60637, USA}

\begin{abstract}
We demonstrate a topological classification of vortices in three dimensional time-reversal invariant topological superconductors based on superconducting Dirac semimetals with an s-wave superconducting order parameter by means of a pair of numbers $(N_\Phi,N)$, accounting how many units $N_\Phi$ of magnetic fluxes $hc/4e$ and how many $N$ chiral Majorana modes the vortex carries. From these quantities, we introduce a topological invariant which further classifies the properties of such vortices under linking processes. While such processes are known to be related to instanton processes in a field theoretic description, we demonstrate here that they are, in fact, also equivalent to the fractional Josephson effect on junctions based at the edges of quantum spin Hall systems. This allows one to consider microscopically the effects of interactions in the linking problem. We therefore demonstrate that associated to links between vortices, one has the exchange of quasi-particles, either Majorana zero-modes or $e/2$ quasi-particles, which allows for a topological classification of vortices in these systems, seen to be $\mathbb{Z}_8$ classified. While $N_\Phi$ and $N$ are shown to be both even or odd in the weakly-interacting limit, in the strongly interacting scenario one loosens this constraint. In this case, one may have further fractionalization possibilities for the vortices, whose excitations are described by $SO(3)_3$-like conformal field theories with quasi-particle exchanges of more exotic types.

\end{abstract}

\maketitle 

\section{Introduction}\label{sec:intro}

Under periodic boundary conditions, superconducting vortices can be thought of as closed (quasi) one-dimensional strings immersed in the superfluid matter. From the mathematical point of view, such un-tangled strings are called uknots, and distinguishing the unknot from other classes of tangles, or links between tangles of closed one-dimensional strings is the subject of knot theory \cite{knotbook}.

Strings in superconducting matter, however, distinguish themselves from the original mathematical problem of knot theory. Their quasi-one dimensional nature, together with the vanishing of the superconducting order parameter inside the defect, allows for the existence of inner structure. In general, superconducting vortices are accompanied by bound quantized electronic states (Caroli-de Gennes-Matricon modes) \cite{CdG}, as well as by quantized units of magnetic flux, in contrast to the featureless strings of mathematical knots.

While complicated knotting of superconducting vortices is strongly unfavored energetically, their inner structure can rearrange and respond under mutual linking due to Berry phases build-up, even for simple single links (simple crossings of non-parallel vortices in open geometries). Such phenomena are, in fact, ubiquitous if the dispersion of the underlying metallic fermions of an s-wave superconductors is linear, as in the case of Dirac semi-metals~\cite{Murakami2007,BurkovBalentsPRB11,burkovBalenstPRL11,WanVishwanathSavrasovPRB11,Ashvin_Weyl_review,RMP}. In this case, the fermionic bound states on strings display a chiral spectrum, with gapless linearly dispersing modes along the vortex. Such chiral modes are always prone to anomaly physics (quantum breaking of classical symmetries), itself related to the aforementioned Berry phases build-ups, and typically manifest in $\theta$-term electromagnetic responses of the fermionic matter. Of notable importance, a non-triviality of the chiral dispersing modes in this superconducting scenario is that the existing particle-hole symmetry implies a Majorana behavior of the chiral modes. This means the chiral modes in this case correspond to real valued fermionic degrees of freedom, and as such the chiral modes are neutral particles. The immediate consequence of this is that such modes are insensitive to electromagnetic probing, and whose anomalous behavior is therefore harder to understand from standard field theoretic analyses involving $\theta$-term responses.

It is the goal of this paper to discuss the inner structure rearranging under mutual linking of string-defects in superconducting Dirac semi-metals (Fig. \ref{CdGspec} top, for an illustration). We show that in a superconducting Dirac metal, one can create several distinct types of vortices, which can be classified according to their inner structure. Concretely, the most basic vortices, i.e. the ones which can be used as building blocks for any other complex case, consist of (i) regular/trivial vortices, which carry quantized units of ($\pi$ in natural units) magnetic flux (ii) Dirac strings, which carry a pair of chiral dispersing Majorana modes, and (iii) the so called chiral vortices\cite{QWZ}, which carry a half-unit of magnetic flux together with a single chiral dispersing Majorana mode. We use these to show that linkings of different combinations of these can lead to the exchange  of several types of quasiparticles between them. The exchange possibilities include Majorana zero-modes (from twisting boundary conditions) and $e/2$ fractionally charged quasi-particles. 

\begin{table}[t!]
\begin{tabular}{lll}
$\delta L(\mathcal{C}_1,\mathcal{C}_2)$&QP exchange\\\hline
half-integer&fractional twisted boundary condition\\
odd&Majorana zero mode\\
$\pm2$ (mod 8)& charge $\pm e/2$ QP (mod $2e$)\\
$4$ (mod 8)& charge $e$ QP (mod $2e$)\\
$8$&$2e$ Cooper pair
\end{tabular}
\caption{Summary of quasiparticle (QP) exhange during a linking process between strings $\mathcal{C}_1$ and $\mathcal{C}_2$ with change of linking invariant $\delta L(\mathcal{C}_1,\mathcal{C}_2)$ (see \eqref{lkinv1}).}
\label{tab:summar_intro}
\end{table}

Introducing strong interactions allows us to extend the picture described above. In this case, more exotic vortices are shown to be possible, out of phases with long-range topological order allowed in the superconductor Majorana surfaces. Thus, under strong interactions, we also argue in favor of further fractionalizations of quasi-particle exchanges based on  identifying vortices whose internal matter is shown to be described by the relative tensor product $\mathcal{G}=SO(3)_3\boxtimes\overline{SO(4)_1}$ conformal field theories (CFTs)~\cite{LukaszChenVishwanath,SahooZhangTeo15}. These consist of CFTs with $c=1/4$ central charge, described by counter propagating $SO(3)_3$ and  $SO(4)_1$ Wess-Zumino-Witten theories under a particular process of condensation~\cite{BaisSlingerlandCondensation}, as described in the main text. In  this case, the exchanges we can see involve the possibility of exchange of non-Abelian quasi-particles which can be interpreted as fractions of a Majorana zero-mode. A summary of these described results is presented in table \ref{tab:summar_intro}, as reference for the reader.

From a topological point of view, a loop linking phenomenon is not a smooth phenomenon, as it necessarily involves an overlapping situation between the vortices, which is responsible for allowing quasiparticle exchange. Correspondingly an integer invariant (linking number) may be defined characterizing the linking profile. To describe the exchange of quasi-particles in such linking process, we adopt two complimentary approaches, both avoiding the anomaly analysis and the subtleties of the charge neutrality of the vortex bound states. 

The first approach involves putting the linking number integer invariant (denoted by $lk$ in the main text) together with the topological indices which classify the protected chiral modes inside a vortex (denoted by $(N_\Phi,N)$) in order to define a new linking invariant (denoted by $L(\mathcal{C}^1, \mathcal{C}^2)$)
that can be used to classify the exchanged quasiparticles. Throughout the paper we will be always comparing our results with this invariant, showing the consistency of its definition. 

The second approach is also novel and involves showing that links of string-defects in superconducting Dirac semi-metals can be mapped into the problem of Josephson junctions at the edge of quantum spin-Hall insulators. The problem can then be treated  smoothly, with the linking process being parametrized by the superconducting phase difference across the junction. This way the quasi-particle exchange between linking vortices is treated in terms of topological charge pumps in the Josephson junction. This approach has also the advantage that it allows us also to treat the effects of interactions between particles in the loops. These are shown to affect the quasi-particle exchanges associated with the linking be means of the $\mathbb{Z}_4$ fractional Josephson effect\cite{ZK_PhysRevLett.113.036401}, leading to the mentioned possibility of exchange of $e/2$ charged quasi-particles between the strings. 

These methods are complimentary to the instanton method used in terms of the effective field theory with $\theta$-terms\cite{QWZ,GuQi}, and avoids the necessity of discussing the anomalous properties of chiral Majorana modes.

The paper is organized as follows. In Sec. \ref{sec:strings} we discuss vortices in s-wave superconducting Dirac semi-metals; we show that due to the fermionic internal structure, several possibilities of line-defects can be introduced, and we classify such defects by the number of chiral modes and magnetic fluxes they carry (Appendix \ref{app:appendix} displays an extensive analysis, analytical and numerical, of these defects and their chiral spectrum). In particular, in subsection \ref{sec:linking} we propose a topological quantity (linking invariant) to classify the distinct linking profiles between vortices. In Sec. \ref{sec:halfpump} we discuss how our vortex linking problem can be mapped into a (long) Josephson junction problem at the edge of a quantum spin-Hall insulator. Appendix \ref{app:QSHJJapp} provides a in-depth discussion of this junction problem, extending the previous analysis of Ref. [\onlinecite{ZK_PhysRevLett.113.036401}], and providing the results important for our analysis. Following, Sec. \ref{sec:Majpump} contains examples of the quasi-particle exchange under vortex linking processes, showing how they should be analyzed from the point of view of our linking invariant and the Josephson junction picture. Finally, Sec. \ref{sec:fractionalize} discusses results regarding further possibilities of fractionalization under strong interactions (the Appendix \ref{app:SONWZW} contains some useful details about $SO(N)_k$ current algebras). We conclude in Sec. \ref{sec:conc}.

\section{Vortices and gapless strings}\label{sec:strings}

\subsection{Vortices in Superconducting Dirac Matter}

\begin{figure}[t!]

\subfloat{%
  \includegraphics[width=0.4\columnwidth]{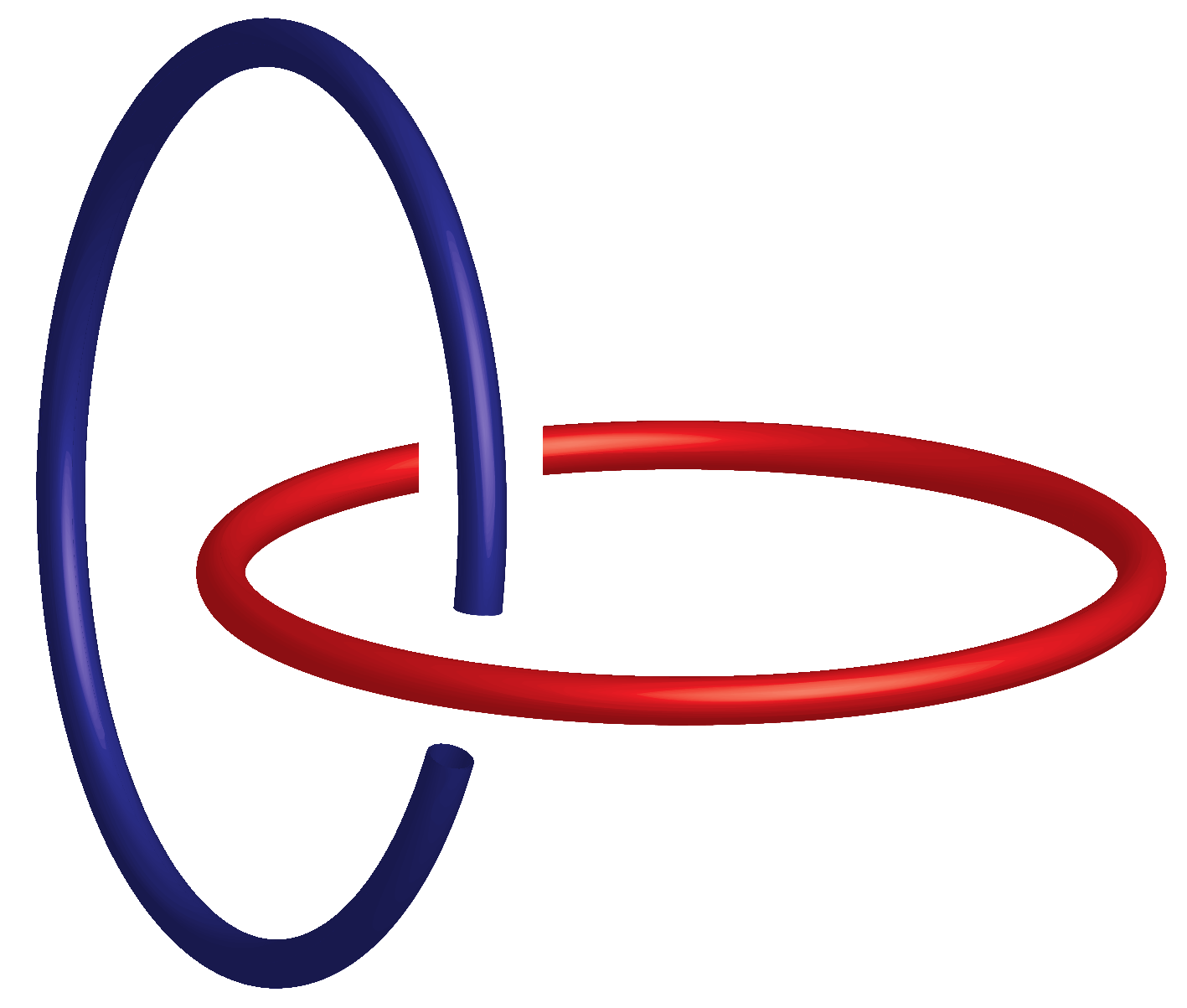}%
}

\subfloat{%
  \includegraphics[width=0.9\columnwidth]{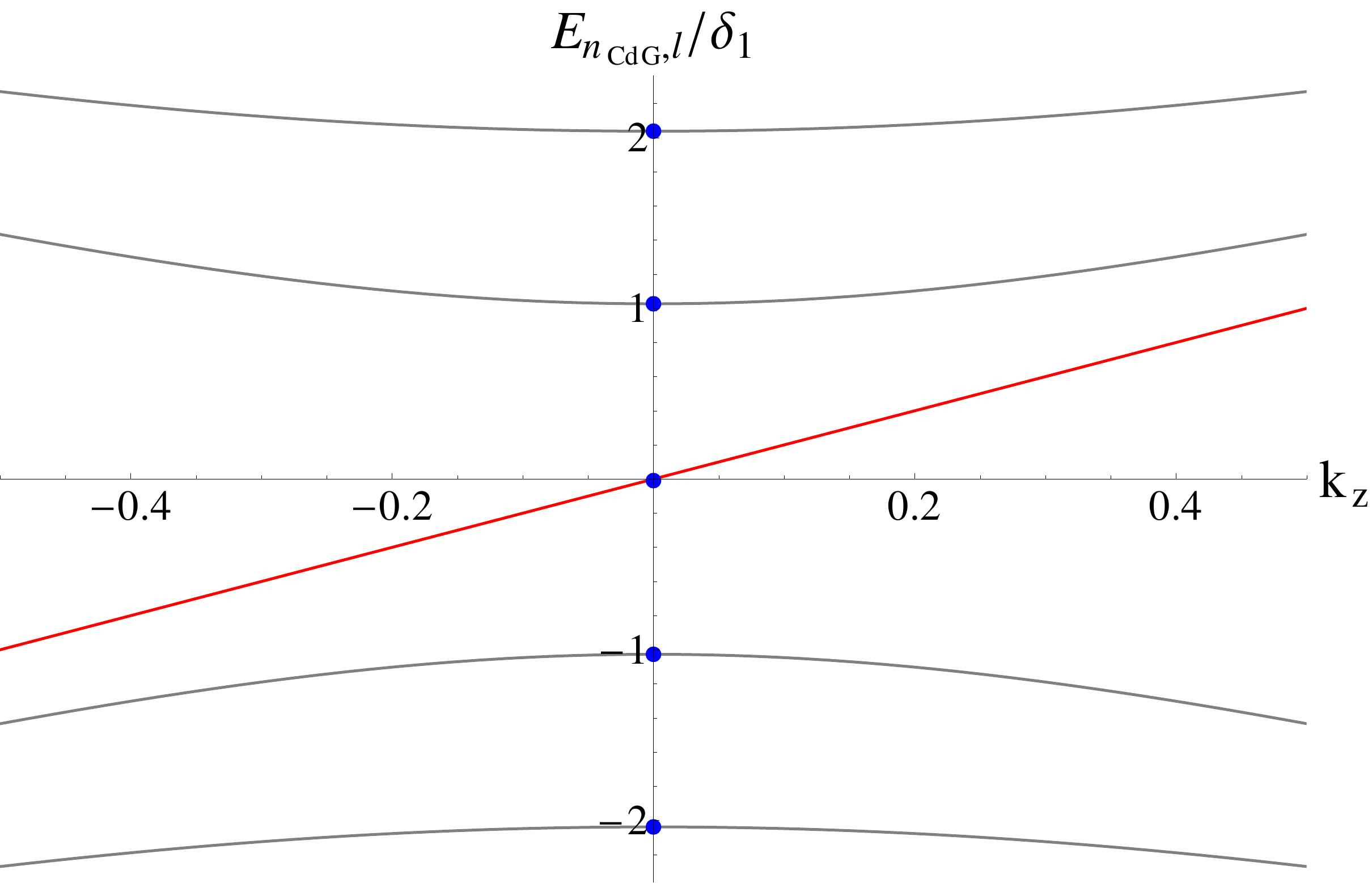}%
}

\caption{(top) A depiction of two linked strings; for us, these are to be seen as vortices in s-wave superconducting Dirac semimetals under periodic boundary conditions. The blue and red loops may represent similar or distinct vortices each carrying a set of $N_\Phi$ flux quanta of $hc/2e$ and of Majorana chiral modes. (bottom)The spectrum of the Caroli-de Gennes bound states at a vortex of an s-wave superconducting Weyl fermion, computed from the method described in the Appendix \ref{app:appendix}. Blue dots correspond to the energies at zero $k_z$ momentum in units of $\delta_1=\Delta_0^2/E_F$, where $\Delta_0$ is the bulk superconducting gap and $E_F$ is the Fermi energy. Of notice is the chiral branch with linear dispersion crossing zero energy, in red. It corresponding to an electromagnetically inert Majorana fermion, due to the Nambu particle-hole constraint of the spinors in the superconductor.} \label{CdGspec}

\end{figure}

There are two classes of topological line defects in three dimensional BCS superconductors. The first type is given by magnetic flux vortices in type II superconductors. Magnetic fields are concentrated along normal tubular regions, and the magnetic flux is quantized in fractional half-units of the magnetic flux quantum $\Phi_0=hc/e$ (or $\Phi_0=2\pi$ when $\hbar=c=e=1$). For a typical $s$-wave SC, the flux vortex is quantized in $\Phi_0/2$ and implies a spatial modulation pairing order parameter $\Delta=\langle c^\dagger_\uparrow c^\dagger_\downarrow\rangle=|\Delta|e^{i\varphi}$, whose complex phase $\varphi$ winds by $2\pi$ around the vortex line. For a more general SC, flux vortices can be further fractionalized in units of $\Phi_0/4$ due to additional order parameters. For example, they appear in $p_x+ip_y$ SC where a half vortex is accompanied by a $\pi$-disclination of the $d$-vector~\cite{SalomaaVolovik85,DasSarmaNayakTewari06,ChungBluhmKim07,Jang_SrRuO}. They can also be present at pair density wave half dislocations in a FFLO-type SC~\cite{ff, lo, radz, radz2, berg09, agterberg08, agterberg11}. Finer fractionalization can occur at lattice disclinations of topological crystalline superconductors~\cite{BenalcazarTeoHughes14,RMP}. In this paper, we focus on strings in superconducting Dirac semimetals where the smallest non-trivial flux vortex carries a magnetic flux of $\Phi_0/4$, as discussed below.

The second type of topological line defects carries gapless electronic degrees of freedom. In an energy scale less than the bulk excitation gap, they are described by $(1+1)$ dimensional fermions and can be chiral or helical, Majorana or Dirac. In the absence of additional symmetries like $U(1)$ charge conservation or time reversal, only chiral modes are protected. They are characterized by a topological quantity known as the chiral central charge $c_-=c_R-c_L=N/2$, which counts the imbalance between right and left moving thermal current at low energies. The $(1+1)$D chiral string can be effectively described in low energy by the Majorana fermion theory (c.f. Appendix \ref{app:appendix})
\begin{align}
\mathcal{L}=\sum_{a=1}^{N_R}i\gamma^a_R(\partial_t-v\partial_x)\gamma^a_R+\sum_{b=1}^{N_L}i\gamma^b_L(\partial_t+v\partial_x)\gamma^b_L 
\end{align} 
where $N=N_R-N_L$, and $\gamma_R^a$, $\gamma_L^b$ are real fermions. Chiral Majorana's are not restricted to vortices, and also arise, for instance, along edges of chiral $p_x+ip_y$ SC, dislocations in weak topological SC, chiral vortices in strong topological SC, and topological insulator -- magnet -- superconductor heterostructures. 

The number $N$ of chiral Majorana species has a topological origin. It is the topological index of the Bogoliubov de Gennes (BdG) defect Hamiltonian $H({\bf k},{\bf r})$ that describes the bulk BdG excitations far away from the defect line~\cite{TeoKaneDefects},
\begin{align}N=\frac{1}{8\pi^2}\int_{\mathrm{BZ}^3\times S^1}\mathrm{Tr}\left(\mathcal{F}_{{\bf k},{\bf r}(\phi)}\wedge\mathcal{F}_{{\bf k},{\bf r}(\phi)}\right),\label{2ndChern}
\end{align}
 where $\phi$ is the spatial angle parameter on a circle $S^1$ that wraps around the line defect, ${\bf k}$ is the three dimensional lattice momentum that lives on the Brillouin zone $\mathrm{BZ}^3$, and $\mathcal{A}_{{\bf k},{\bf r}}^{mn}=\langle u^m_{{\bf k},{\bf r}}|du^n_{{\bf k},{\bf r}}\rangle$ and $\mathcal{F}=d\mathcal{A}+\mathcal{A}\wedge\mathcal{A}$ are the Berry connection and curvature derived from the occupied states $u^m_{{\bf k},{\bf r}}$ of the defect Hamiltonian $H({\bf k},{\bf r})$ . 
 
 As an example, consider the BdG Hamiltonian of an s-wave superconducting Weyl fermion (or a Weyl fermion with a Majorana mass). It takes the form  
\begin{align}H_{\mathrm{Weyl-SC}}({\bf k})=\rho_z\left(\hbar v{\bf k}\cdot{\vec \sigma}-\mu\right)+\Delta_x\rho_x+\Delta_y\rho_y. \label{eq:WeylSCH}
\end{align}
With $\xi_{\bf k}=(c_{\uparrow{\bf k}},c_{\downarrow{\bf k}},c_{\downarrow-{\bf k}}^\dagger,-c_{\uparrow-{\bf k}}^\dagger)$ as the Nambu vector, $\Xi=\rho_y \sigma_yK$ is the particle-hole symmetry operator and $\vec s$ and $\vec\rho$ are Pauli matrices acting on the spin and particle-hole grading. In the presence of a vortex, the pairing phase of the order parameter $\Delta(\phi)=\Delta_x+i\Delta_y=|\Delta_0|e^{i\varphi(\phi)}$ winds by $2\pi N$ around the vortex line. The second Chern invariant \eqref{2ndChern} can be simplified to the winding number $N=\frac{1}{2\pi i}\int_0^{2\pi}d\phi e^{-i\varphi(\phi)}\partial_\phi e^{i\varphi(\phi)}$ and the vortex holds $N$ chiral Majorana fermions. For instance, when $N$=1, the defect Hamiltonian $H_{\mathrm{Weyl-SC}}({\bf k},\phi)$ is structurally identical to the one that describes a time reversal breaking domain wall on the surface of a topological class DIII SC~\cite{SchnyderRyuFurusakiLudwig08,Kitaevtable08,QiHughesRaghuZhang09,HasanKane10,QiZhangreview11,RMP}. In this case, $\Delta_x$ changes sign across the topological SC to vacuum surface interface, and $\Delta_y$ changes sign along the surface between the adjacent domains with opposite time reversal breaking orientations. The line defect sandwiched between the two domains carries a single chiral Majorana fermion. To illustrate, the chiral modes can also be found as reminiscent of the standard (gapped) Caroli-de-Gennes-Matricon modes of superconducting vortices. The chiral spectrum of the vortex bound states can then be constructed explicitly, following the discussion in Appendix \ref{app:appendix}, and is displayed in Fig. \ref{CdGspec}.

In more general grounds, topological strings in three dimensional superconductors are composites of flux vortices and chiral Majorana modes. They are characterized by two integers $(N_\Phi,N)$. The first corresponds to the magnetic flux $\Phi=N_\Phi\Phi_0/4$ running along the string. The second counts the number of chiral Majorana fermions or the chiral central charge $c_-=N/2$. The vortex in a superconducting Weyl fermion as described above is an example with $(N_\Phi,N)=(2N,N)$. 

Unfortunately, a single Weyl fermion violates the fermion doubling theorem and cannot exist by itself in a true non-holographic three dimensional system. Instead, we focus on a massless Dirac fermion, that is, two Weyl fermions with opposite chiralities, described by the Bloch Hamiltonian 
\begin{align}
H_{\mathrm{Dirac}}({\bf k})=\hbar v{\bf k}\cdot{\tau_z \vec \sigma}
\end{align} 
where now $\vec{\tau}$ Pauli matrices act on some orbital or sublattice degrees of freedom, characterizing the chiralities. The BdG Hamiltonian of a superconducting \emph{Dirac} fermion then takes the form 
\begin{align}
H_{\mathrm{Dirac-SC}}({\bf k})=\rho_z \left(\hbar v{\bf k}\cdot{\tau_z \vec \sigma}-\mu\right)+\Delta\Gamma\label{Dirac-SC}
\end{align}
 where $\mu$ is the Fermi energy, the Nambu vector is now chosen to include the $\tau$ degrees of freedom as $\xi_{\bf k}=(c_{{\bf k}\tau \sigma},i(\sigma_y)_{\sigma \sigma'}c^\dagger_{-{\bf k}\tau \sigma'})$, and $\Delta\Gamma$ is the off-diagonal superconductor pairing order parameter. 
 
Restricting ourselves to s-wave superconductivity, fermion statistics dictates that there are 10 linearly independent $s$-wave pairing terms $\rho_{x,y},\rho_{x,y}\tau_z,\rho_{x,y}\tau_y{\vec \sigma}$. We are interested in pairings that gap the Dirac semimetal, so that we can neglect the last 6 terms above and assume the restricted linear combination 
\begin{equation}
\Delta\Gamma=\Delta_x\rho_x+\Delta_y\rho_y+\Delta^\tau_x\rho_x\tau_z+\Delta^\tau_y\rho_y\tau_z.
\end{equation} 
Each term in $\Delta\Gamma$ now anticommutes with $\rho_z(H_{\text{Dirac}}-\mu)$ and gives rise to an energy gap $E_{\text{gap}}=\text{min}\left| \Delta\pm\Delta^\tau \right|$, for $\Delta=\Delta_x+i\Delta_y$ and $\Delta^\tau=\Delta^\tau_x+i\Delta^\tau_y$. Written in second quantized form, the $s$-wave pairing is $\boldsymbol\Delta_{\tau\sigma \tau' \sigma'}c^\dagger_{-{\boldsymbol{k}}\tau\sigma}c^\dagger_{{\boldsymbol{k}}\tau' \sigma'}+h.c.$ where 
\begin{align}
\boldsymbol\Delta_{\tau\sigma \tau' \sigma'}=i(\sigma_y)_{\sigma' \sigma}\left(\delta_{\tau\tau'}\Delta+(\tau_z)_{\tau\tau'}\Delta^\tau\right).
\end{align} 
The magnetic flux along a vortex string $\Phi=N_\Phi\Phi_0/4=N_\Phi hc/4e$ is identical to the winding number of the pairing matrix \begin{align}N_\Phi=\frac{1}{2\pi i}\int_0^{2\pi}\frac{1}{2}\mathrm{Tr}\left[\boldsymbol\Delta(\phi)^{-1}\partial_\phi\boldsymbol\Delta(\phi)\right]d\phi\label{Nphi}\end{align} where $\phi$ is the polar angle about the string, and the $1/2$ factor comes from the two orbital/sublattice species $\tau=\pm1$.  

We define the sum and difference $\Delta_\pm=\Delta\pm\Delta^\tau$, which are the pairing order parameter in the $\tau_z=\pm1$ sectors. The parameter space for the 2-component pairing $(\Delta_+,\Delta_-)=(|\Delta_+|e^{i\varphi_+},|\Delta_-|e^{i\varphi_-})$, when the pairing gap $E_{\mathrm{gap}}=\min|\Delta_\pm|$ is non-vanishing, is topologically equivalent (by a strong deformation retract) to the 2-torus $T^2=S^1\times S^1$ parametrized by the two pairing phases $\varphi_\pm$. The fundamental group $\pi_1(T^2)=\mathbb{Z}\times\mathbb{Z}$ has two generators corresponding to the winding of the two pairing phases, and therefore there are two primitive superconducting vortices, one winds $\varphi_+$ by $2\pi$ around the vortex and the other winds $\varphi_-$ by $2\pi$. A vortex in general is a combination $\Delta_\pm=|\Delta_\pm|e^{2\pi im_\pm\phi}$, for $\phi$ the polar angle about the vortex line, and is characterized by the two winding numbers $(m_+,m_-)$.

Vortex lines can equivalently be characterized by its magnetic flux and the chiral Majorana fermion it carries. An $hc/2e$ flux vortex corresponds to $2\pi$ windings in both $\varphi_\pm$, i.e.~$(m_+,m_-)=(1,1)$. On the other hand, when $m_+=-m_-$, there is no net magnetic field running long the vortex. In general, a $(m_+,m_-)$ vortex thus carries a magnetic flux $\Phi=(m_++m_-)hc/4e$, or $N_\Phi=m_++m_-$. Furthermore, as $\sigma_z$ is an (artificial) symmetry in the BdG Hamiltonian \eqref{Dirac-SC}, and $\sigma_z=\pm1$ represents the two Weyl fermions with opposite chirality, a $(m_+,m_-)$ vortex carries $m_+$ (or $m_-$) chiral Majorana fermion in the $+$ (resp.~$-$) sector. Together, it has the net Majorana chirality $N=m_+-m_-$. Hence, topological strings in superconducting Dirac semimetals are characterized by $(N_\Phi,N)$, where $N_\Phi$ and $N$ are integers with the same parity. Table \ref{tab:nomenc} summarizes some nomenclature and particularly relevant topological strings for us.
\begin{table}
\begin{centering}
\begin{tabular}{|c|c|c|}
\hline 
$\left(N_{\Phi},N\right)$ & $\left(m_+,m_-\right)$ & nomenclature\tabularnewline
\hline 
\hline 
$\left(2,0\right)$ & $\left(1,1\right)$ & trivial vortex\tabularnewline
\hline 
$\left(0,2\right)$ & $\left(1,-1\right)$ & Dirac string\tabularnewline
\hline 
$\left(1,1\right)$ & $\left(2,0\right)$ & chiral vortex\tabularnewline
\hline 
\end{tabular}
\par\end{centering}

\caption{Some particular topological strings whose importance gives them rights to a nomenclature
of their own. The first case is the trivial vortex, which carries
a $hc/2e$ ($\pi$) flux and has no internal structure. This
is the same object that exists in standard, non-topological, type
II s-wave superconductors. Second is the Dirac vortex, which has $N=2$,
carrying a pair of Majorana chiral fermions, or a regular and charge
chiral fermion. It has opposite windings in $\Delta_{+}$ and $\Delta_{-}$,
which leads to a zero net-flux. Finally, the chiral vortex carries
a single Majorana chiral mode. The total magnetic flux bound to it
is $\Phi_{0}/4$ ($\pi/2$) \cite{GuQi} \label{tab:nomenc}}
\end{table}

Lastly, we notice that the $(N_\Phi,N)$ characterization extends to general $s$-wave pairing, which allows $\rho_{x,y}\tau_y {\vec \sigma}$ terms in $\Delta\Gamma$ as long as it gives a non-vanishing pairing gap. This is because $(N_\Phi,N)$ corresponds to physical quantities \eqref{Nphi} and \eqref{2ndChern}, namely magnetic flux and chiral thermal current, which are well-defined in the general setting (the antisymmetry of the pairing matrix $\boldsymbol\Delta=-\boldsymbol\Delta^T$ in general guarantees the normalized winding number \eqref{Nphi} to be integral.) The restriction that $N_\Phi$ and $N$ are either both even or both odd can be relaxed by introducing additional long-range topological order in the presence of strong many-body interactions. This will not be discussed until the later part of this paper in section~\ref{sec:fractionalize} where we discuss fractional vortices.

\subsection{Linking invariant}\label{sec:linking}
Two disjoint closed topological strings in three dimensions can link. In particular, we are interested in a linking invariant  which we define as
\begin{equation}
L(\mathcal{C}^1,\mathcal{C}^2)\equiv \frac{\left(N_\Phi^{(1)}N^{(2)}+N_\Phi^{(2)}N^{(1)}\right)}{2} lk(\mathcal{C}^1,\mathcal{C}^2).\label{lkinv1}
\end{equation}
Here, $\mathcal{C}^1$ and $\mathcal{C}^2$ are strings with corresponding magnetic fluxes $\Phi^{(i)}=N_\Phi^{(i)}\Phi_0/4$ and chiral Majoranas $\psi_a^{(i)}$, $a=1,\ldots,N^{(i)}$, with $i=1,2$, and where $lk(\mathcal{C}^1,\mathcal{C}^2)$ is the integral linking number between the two loops, whose definition we now explain.

The magnetic field along string $\mathcal{C}^1$ is $B^{(1)}({\bf r})=(N_\Phi^{(1)}\Phi_0/4)\delta_{\mathcal{C}^1}({\bf r})$. Here $\delta_{\mathcal{C}^1}$ is the Dirac delta 2-form, in the limit when the London penetration depth is infinitesimal, so that the integral $\int_\Sigma\delta_{\mathcal{C}^1}=|\Sigma\cap\mathcal{C}^1|$ over any (open) surface $\Sigma$ in real 3-space is identical to the number of intersection points, counting multiplicities and orientation, between $\Sigma$ and $\mathcal{C}^1$. Let $A^{(1)}({\bf r})$ be a vector potential, $dA^{(1)}=B^{(1)}$. For example, using an open surface $\mathcal{S}^1$ whose boundary is $\partial\mathcal{S}^1=\mathcal{C}^1$, the vector potential can be chosen to be $A^{(1)}({\bf r})=(N_\Phi^{(1)}\Phi_0/4)\delta_{\mathcal{S}^1}({\bf r})$, where $\delta_{\mathcal{S}^1}$ is the Dirac delta 1-form so that the integral $\oint_{\mathcal{P}}\delta_{\mathcal{S}^1}=|\mathcal{P}\cap\mathcal{S}^1|$ over any closed loop $\mathcal{P}$ in real 3-space is the intersection 
number (counting multiplicities and orientation) between $\mathcal{P}$ and $\mathcal{S}^1$. For instance, $d\delta_{\mathcal{S}^1}=\delta_{\partial\mathcal{S}^1}=\delta_{\mathcal{C}^1}$, and the linking number between $\mathcal{C}^1$ and $\mathcal{C}^2$ is the intersection number \begin{align}lk(\mathcal{C}^1,\mathcal{C}^2)=\oint_{\mathcal{C}^2}\delta_{\mathcal{S}^1}=\int_{\mathbb{R}^3}\delta_{\mathcal{C}^2}\wedge\delta_{\mathcal{S}^1}.\label{lknum}\end{align}

The linking invariant can be re-expressed out of physical quantities only. Notice that the chiral Majorana fermions along $\mathcal{C}^i$ is topologically protected by the second Chern invariant \eqref{2ndChern}. One can define a closed differential 1-form by a momentum space integral \begin{align}
\alpha^{(i)}({\bf r})=\frac{1}{8\pi^2}\int_{\mathrm{BZ}^3}\mathrm{Tr}\left(\mathcal{F}^{(i)}_{{\bf k},{\bf r}}\wedge\mathcal{F}^{(i)}_{{\bf k},{\bf r}}\right). 
\end{align}
Here the Berry curvature $\mathcal{F}^{(i)}_{{\bf k},{\bf r}}$ is defined by the occupied states of the defect Hamiltonian $H^{(i)}({\bf k},{\bf r})$ generated {\em only} by the defect string $\mathcal{C}^i$. Then, the closed 1-form for a given string $i$, $\alpha^{(i)}({\bf r})$ is still well-defined at positions ${\bf r}$ far away from $\mathcal{C}^i$, in particular, along another string $\mathcal{C}^j$. As $\int_{\mathcal{C}^j}\alpha^{(i)}=N^{(i)}lk(\mathcal{C}^i,\mathcal{C}^j)$, the linking invariant \eqref{lkinv1} reads 
\begin{align}
L(\mathcal{C}^1,\mathcal{C}^2)&=
\frac{N_\Phi^{(1)}\int_{\mathcal{C}^1}\alpha^{(2)}+N_\Phi^{(2)}\int_{\mathcal{C}^2}\alpha^{(1)}}{2}\nonumber \\
&=\frac{2}{\Phi_0}\int_{\mathbb{R}^3}\left[B^{(1)}\wedge\alpha^{(2)}+B^{(2)}\wedge\alpha^{(1)}\right]\nonumber\\
&=\frac{\sum_{i \neq j}}{(2\pi)^2\Phi_0}\int_{\text{BZ}^3\times\mathbb{R}^3}   B^{(i)}({\bf r})\wedge\mathrm{Tr}\left(\mathcal{F}^{(j)}_{{\bf k},{\bf r}}\wedge\mathcal{F}^{(j)}_{{\bf k},{\bf r}}\right).\label{lkinv2}
\end{align}

The invariant is stable against perturbation and deformation. It cannot change unless the bulk gap closes or when topological strings cross. The linking invariant is additive in the sense that \begin{align}L(\mathcal{C}^1\cup\mathcal{C}^2,\mathcal{C}^3)&=L(\mathcal{C}^1,\mathcal{C}^3)+L(\mathcal{C}^2,\mathcal{C}^3)\nonumber\\L(\mathcal{C}^1,\mathcal{C}^2\cup\mathcal{C}^3)&=L(\mathcal{C}^1,\mathcal{C}^2)+L(\mathcal{C}^1,\mathcal{C}^3).\end{align} This is natural because, as a consequence of the conservation of energy and magnetic field $dB=0$, the magnetic flux and Majorana chirality along topological strings are also additive.

From now on, we will denote an string as $\mathcal{C}_{(N_\Phi,N)}$, identifying its structural content, and a given linking process as $\mathrm{L}_{lk}(\mathcal{C}^1_{(N^1_\Phi,N^1)},\mathcal{C}^2_{(N^2_\Phi,N^2)})$, identifying each of the strings participating in the process, as well as the value of the linking number $lk$. These informations are enough to, from the topological point of view, uniquely characterize the linking invariant and the given linking process.

\section{Chiral strings linking and quantum pumping}\label{sec:halfpump}

The process of linking pairs of strings may lead to a change in the structure of the ground state of the string modes. We devote this Section to the development of an analysis capturing such changes in terms of a quantum pump language and the arising topological consequences. 

We approach the problem from a simple non-trivial linking process between string-objects, the linking between two chiral vortices (i.e. type $(1,1)$). This is represented by the linking invariant $\mathrm{L}_{lk}(\mathcal{C}^1_{(1,1)},\mathcal{C}^2_{(1,1)})$. As a first example of application of our  invariant, notice that a  \emph{double} linking of two chiral vortices $(1,1)$ or a \emph{single} linking between a trivial vortex $(2,0)$ with a Dirac string $(0,2)$  return the same value: $\mathrm{L}_{1}(\mathcal{C}^1_{(0,2)},\mathcal{C}^2_{(2,0)})=\mathrm{L}_{2}(\mathcal{C}^1_{(1,1)},\mathcal{C}^2_{(1,1)})=2$; such processes are thus topologically equivalent and the results of this section can also be constructed in terms of the latter, likewise as in terms of the former.

In order to study how the ground state evolution develops in this scenario, we demonstrate here a mapping in which the linking event can be described via an adiabatic process. Through our mapping, we see that string linking processes in Dirac semimetals are equivalent to cyclic phase evolutions across Josephson junctions at the edges of quantum spin-Hall phases; via this picture, the linking process can be studied in full detail, including also effects of fermionic interactions. The process we use here does not require the system's action and is complimentary to the effective field theory approach with instanton processes describing vortex crossings, avoiding the drawbacks of having to consider anomalous electromagnetic responses of neutral fermions\cite{QWZ, GuQi, StoneLopes}.

We start by showing the details of our mapping between chiral vortices linking and quantum spin-Hall Josephson junctions, then reviewing the physics of ground state evolution of such particular junctions in the presence of interactions and in the very-long-junction limit, which is our case; finally we display the consequences for physics in some examples.

\subsection{Vortex Linking-QSHJJ mapping}

We start considering a simple scenario with a pair of strings, each with a single chiral Majorana mode. Closing the strings into loops, their Hamiltonian reads
\begin{eqnarray}
H_{links}&=&\frac{v}{2\pi R}\int_{0}^{2\pi}d\theta\left(\gamma^{1},\gamma^{2}\right)\left(\begin{array}{cc} 
-i\partial_{\theta} \\
 & -i\partial_{\theta}
\end{array}\right)\left(\begin{array}{c}
\gamma^{1} \nonumber \\
\gamma^{2}
\end{array}\right)\\&=&\frac{v}{R}\sum_{l}\left(\gamma^{1}_{-l},\gamma^{2}_{-l}\right)\left(\begin{array}{cc}
l\\
 & l
\end{array}\right)\left(\begin{array}{c}
\gamma^{1}_{l}\\
\gamma^{2}_{l}
\end{array}\right), 
\end{eqnarray}
where $1,2$ label each loop, $\gamma^i_l$ is a real fermion living on the loop with angular momentum $l$, and $v$ and $R$ are a velocity scale and loop radius. $l$ is an integer or half-integer, depending on the boundary conditions in the Majorana fermions. Due to the conformal symmetry of the problem, no generality is lost by taking loops of equal radii. Notice that the chiral Majorana fermions of the loops can be combined and this problem can equivalently be described in terms of a charged Dirac chiral mode, connecting with the discussion of the equivalence between linkings of chiral vortices and of regular vortices with Dirac strings.

Still making use of the conformal symmetry, we can perform a trick \cite{Fendley20091547}. We squash the loop, obtaining a line segment; the chiral Majorana modes from each loop become then a pair of counter propagating Majorana chiral modes in a single line. For the new pair of helical Majorana fermions, we can write an action
\begin{eqnarray}
S_{0}&=&v\int d^{2}x\left[i\gamma_{R}^1\partial_{+}\gamma_{R}^1+i\gamma_{L}^1\partial_{-}\gamma_{L}^1\right]
\nonumber \\
&&+v\int d^{2}x\left[i\gamma_{R}^2\partial_{+}\gamma_{R}^2+i\gamma_{L}^2\partial_{-}\gamma_{L}^2\right]
\end{eqnarray} where $\gamma_{R/L}^i$ are now the right and left movers from each line $i=1,2$ and $\partial_{\pm}=v^{-1}\partial_{t}\pm\partial_{x}$. Here the spatial extension of each line segment is taken as $x:0 \rightarrow L= \pi R$.
The mapping between the loop chiral Majorana modes and line-segment helical Majorana modes is not complete if one does not specify the boundary conditions at each end of the line segment, in order to recover the behavior of the full loop theory. To implement such boundary conditions, we augment the action with the following boundary term
\begin{eqnarray}
S&=&S_{0}+\int dtL_{b}, \\
L_{b}&=&-iav\gamma_{L}^1\left(0\right)\gamma_{R}^1\left(0\right)+ibv\gamma_{L}^1\left(L\right)\gamma_{R}^1\left(L\right) \nonumber\\
&&-iav\gamma_{L}^2\left(0\right)\gamma_{R}^2\left(0\right)+ibv\gamma_{L}^2\left(L\right)\gamma_{R}^2\left(L\right).
\end{eqnarray}
To see that this indeed fixes the boundary conditions as desired, one simply compute the equations of motion. For the right movers of each loop, for example, they read
\begin{eqnarray}
2\left(\partial_{0}+\partial_{1}\right)\gamma_{R}^1&=&\delta\left(x-L\right)\left(\gamma_{R}^1-b\gamma_{L}^1\right) \nonumber \\ &&-\delta\left(x\right)\left(\gamma_{R}^1-a\gamma_{L}^1\right),\\
2\left(\partial_{0}+\partial_{1}\right)\gamma_{R}^2&=&\delta\left(x-L\right)\left(\gamma_{R}^2-b\gamma_{L}^2\right) \nonumber \\ &&-\delta\left(x\right)\left(\gamma_{R}^2-a\gamma_{L}^2\right),
\end{eqnarray} 
where boundary contributions appear in integrations by parts. This implies that at the edges
\begin{eqnarray}
\gamma^{1,2}_R\left(0\right)&=&a\gamma^{1,2}_L\left(0\right), \\ 
\gamma^{1,2}_R\left(L\right)&=&b\gamma^{1,2}_L\left(L\right).
\end{eqnarray}
The corresponding equations of motion for the left movers can be obtained by exchanging L and R, so that consistency fixes $a^2=1$ and $b^2=1$. This way, each loop can have a fixed boundary condition, either periodic or anti-periodic, enforced by $ab=1$ or $ab=-1$, respectively.

This result is known~\cite{Fendley20091547}, and general for a pair of disjoint Majorana chiral loops. For linking loops, however, it is not the most general possibility. In the process of loop linking there necessarily exists a regime in which scattering of modes from one loop to the other is activated. For a pair of chiral vortices, it should also be expected that such linking process leads to a twist in boundary conditions. One can understand this from a topological point of view as follows: each chiral vortex carries a $\pi/2$ magnetic flux. Since vortices of opposite chirality have the Majorana modes in disconnected Hilbert spaces, one can link two vortices of the same chirality with a third one of opposite chirality without changing our topological invariant. While one does not expect this to lead to any new scattering of quasi-particles, due to the disconnected Hilbert spaces, one sees that the three joint vortices have a total of $\pi$ flux piercing each loop (i.e. each loop has 2 piercing loops each carrying $\pi/2$ for a total of $\pi$ flux). It is then a known fact that piercing a chiral Majorana loop by a $\pi$-flux leads to a change in its boundary condition~\cite{aliceaRev}. 

To account for the discussion above, we write a generalized boundary Lagrangian
\begin{equation}
L_{b}=-iv\delta\left(x\right)a_{ij}\gamma_{L}^i\gamma_{R}^j+iv\delta\left(x-L\right)b_{ij}\gamma_{L}^i\gamma_{R}^j.
\end{equation}
This way, the boundary conditions can be thought of as a mapping of the spinor $\boldsymbol{\gamma}=\left(\gamma_{R}^1,\gamma_{R}^2,\gamma_{L}^1,\gamma_{L}^2\right)^{T}$ onto itself at $x=0$ or $x=L$. If we take $x=0$, for example, the boundary conditions become
\begin{equation}
\boldsymbol{\gamma}\left(0\right)=A\boldsymbol{\gamma}\left(0\right),
\end{equation} 
where
\begin{eqnarray}
A=\left(\begin{array}{cccc}
 &  & a_{11} & a_{21}\\
 &  & _{12} & a_{22}\\
a_{11} & a_{12}\\
a_{21} & a_{22}
\end{array}\right).
\end{eqnarray}
Now, from $\gamma=A\gamma$, the boundary coupling matrix must square to the identity $A^{2}=\mathbf{I}_{4\times4}$. A conveniently parametrized solution to this is
\begin{eqnarray}
a_{11}&=&-\cos\varphi
\\a_{22}&=&-\cos\varphi
\\a_{12}&=&\sin\varphi
\\a_{21}&=&-\sin\varphi.
\end{eqnarray}
A similar set of equations can be derived for the coefficients $b_{ij}$, but as for the disjoint links, the overall boundary conditions can be fixed by a single set of parameters, which we choose here to be $a_{ij}$ parametrized by the angle $\varphi$, setting $b_{ij}=\delta _{ij}$ for simplicity.
Some sign ambiguity arises when deriving these. The ambiguities can be fixed by noticing that in scenario with two linked loops can be chosen to correspond to $\varphi=0$; both loops should have anti-periodic boundary conditions in this case, allowing to fix the sign ambiguities. Such a choice will prove convenient for the Josephson junction scenario coming below.

In general, the new boundary Lagrangian can thus be written,
\begin{eqnarray}
&&L_{b}=  \nonumber\\
&&iv\left[\cos\varphi\left(\gamma_{L}^1\gamma_{R}^1+\gamma_{L}^2\gamma_{R}^2\right)-\sin\varphi\left(\gamma_{L}^1\gamma_{R}^2-\gamma_{L}^2\gamma_{R}^1\right)\right]\delta\left(x\right) \nonumber \\
&&+iv\left[\left(\gamma_{L}^1\gamma_{R}^1+\gamma_{L}^2\gamma_{R}^2\right)\right]\delta\left(x-L\right)
\end{eqnarray}
The boundary conditions at $x=0$ then follow as
\begin{eqnarray}
\gamma_{R}^1\left(0\right)&=&-\cos\varphi\gamma_{L}^1\left(0\right)-\sin\varphi\gamma_{L}^2\left(0\right) \\
\gamma_{R}^2\left(0\right)&=&\sin\varphi\gamma_{L}^1\left(0\right)-\cos\varphi\gamma_{L}^2\left(0\right).
\end{eqnarray}

\begin{figure}[t!]
\begin{centering}
\includegraphics[width=1. \linewidth]{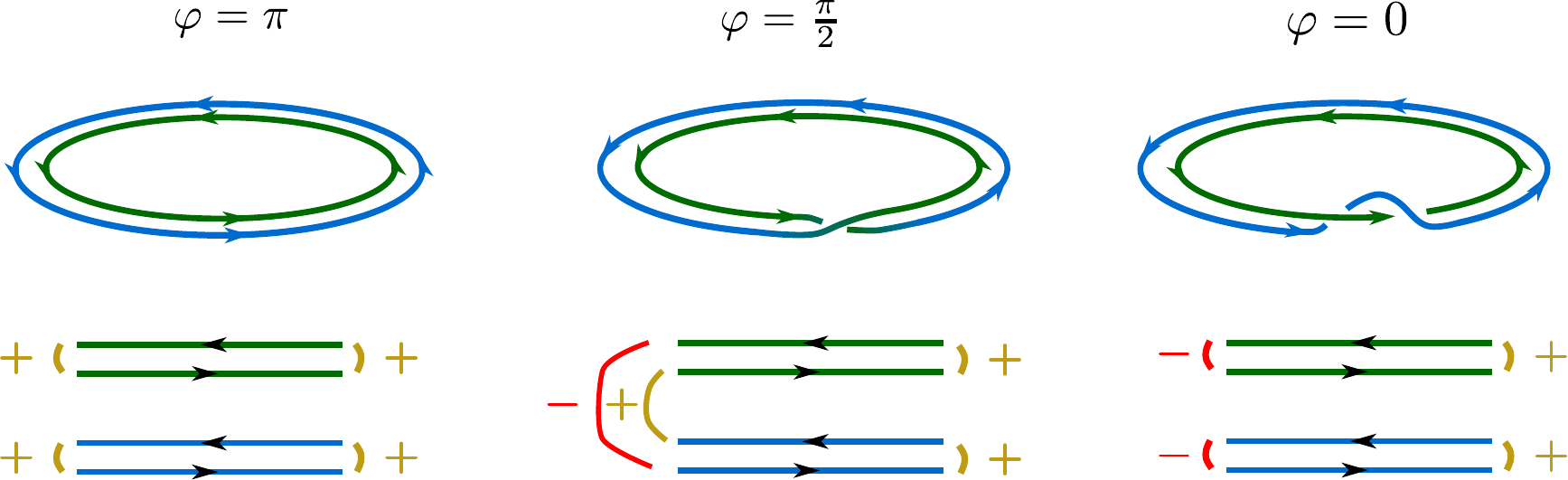}

\par\end{centering}
\protect\caption{Single Majorana chiral loop linking (top) and the squashed line segment picture (bottom). Blue and green correspond to two distinct topological strings carrying  Majorana chiral modes, with same chirality, and a $\Phi_0/4$ ($\pi/2$) magnetic flux. (top) The linking process is parametrized by an angle  $\varphi$ which we display at some special points $\varphi=0i,\pi/2,\,\pi$. These special values correspond to a pair of linked loops, a single crossing over between the loops ( generating a single large loop) and finally to a pair of unlinked loops. (bottom) Squashing the loops generates line segments with non-chiral Majorana states. The loop linking picture is recovered by considering proper signs and mixings of the boundary scatterings for each value of the $\varphi$ evolution parameterizing the linking process.} \label{link_evo}
\end{figure}

Interpreting these equations is a simple matter. At $\varphi=0$, we have a pair of linked Majorana chiral loops, both with anti-periodic boundary conditions. At $\varphi=\pi/2$, the loops have merged into a single large loop, with right movers from loop 1 scattering into left movers of loop 2, and vice versa. Finally, at $\varphi=\pi$, the loops are disjoint again, but the boundary conditions have changed from anti-periodic to periodic in both loops; the process of evolving $\varphi:0\to\pi$ is interpreted as the unlinking of a pair of loops. This is depicted diagrammatically in Fig.\ \ref{link_evo}.

While these results are enough for us, in the sense that we have succeeded in writing a continuous mapping for the evolution of two linking loops, we can proceed one step further. The squashing of loops into line segments artificially double our set of fermionic modes in the bulk of the lines; these can be then combined into complex fermions. Choosing a gauge such that
\begin{align}
&\psi_{R}=\frac{\gamma_{R}^1+i\gamma_{R}^2}{\sqrt{2}},\quad  \psi_{R}^{\dagger}=\frac{\gamma_{R}^1-i\gamma_{R}^2}{\sqrt{2}},
\nonumber \\
& \psi_{L}=\frac{i\gamma_{L}^1+\gamma_{L}^2}{\sqrt{2}},\quad \psi_{L}^{\dagger}=\frac{-i\gamma_{L}^1+\gamma_{L}^2}{\sqrt{2}},
\end{align}
where $\{\psi_{i},\psi_{j}^{\dagger}\} =\delta_{ij}$, the corresponding fermionic Hamiltonian may be written
\begin{eqnarray}
H_{BdG}&=&\int dx\Psi^{\dagger}h\left( \varphi \right)\Psi  \label{eq:JJBdG}\\
&=& \int dx\Psi^{\dagger}\left[-iv\rho_{z}\tau_{z}\partial_{x}+\rho_{x}\Delta_{1}+\rho_{y}\Delta_{2}\right]\Psi, \nonumber
\end{eqnarray}
where $\Psi=(\psi_{R},\psi_{L},\psi_{L}^{\dagger},-\psi_{R}^{\dagger})^{T}$ and $\tau$ and $\sigma$ matrices live in the Hermitian conjugate and left/right mover spaces, respectively. Particularly important to notice is the fact that the boundary terms got mapped into an s-wave superconducting coupling
\begin{equation}
\Delta\left(x\right)=\Delta_{1}+i\Delta_{2}=\frac{v}{2}\left[\delta\left(x-L\right)+e^{i\varphi}\delta\left(x\right)\right].
\end{equation}
This corresponds to a Josephson junction at the edge of a quantum spin-Hall insulator in the extreme long-junction limit \cite{ZK_PhysRevLett.113.036401} (superconductivity is introduced only in point contacts at $x=0$ and $L$). The quasiparticle exchange between two Majorana chiral loops under a linking event can thus be probed by studying the ground state evolution of a corresponding Josephson junction in the edge of a quantum spin-Hall insulator. As shown in Appendix \ref{app:QSHJJapp}, following \cite{ZK_PhysRevLett.113.036401}, the fractional Josephson effect in such junctions pumps quasiparticles with a minimum charge of $e/2$, fixing the periodicity of the Hilbert space under evolution of the superconducting phase difference at $8\pi$, in contrast with the $2\pi$ periodicity of the Hamiltonian. Translating back, this provides a means to fix the quasi-particle exchange possibilities under linking phenomena in the vortex context. In what follows, we provide a discussion of these results, fixing the topological properties of the links, and showing that the analysis based on the Josephson effect match the predictions of our invariant defined in \eqref{lkinv1}.

\section{Notable Examples}\label{sec:Majpump}



We now return to our discussion about quasi-particle exchanges and vortex linking.  Using the crossing between $\mathcal{C}_{(1,1)}$ strings as an example, we have studied the phenomenon of their inner structure rearrangement/quasi-particle exchange in terms of the fractional Josephson effect in junctions at the edes of quantum spin-Hall insulators. This is the most basic linking process that needs to be considered if one wants to build, topologically, the results of linkings between any set of arbitrary string-defects in superconducting Dirac semi-metals. 
In this section, we will discuss the generalities of the chiral vortex linking in the light of our previous results and use it as a building block to consider the other simple examples of links and quasi-particle exchanges, which can be extended for even more complicated scenarios. Overall, we will show that vortices of Dirac semi-metals wish superconducting s-wave pairings can be linked 8 times before recovering the starting point. In the discussions, we will keep close contact with our linking invariant, showing its robustness, and how to use it in order to topologically connect the quasi-particle exchanging results between seemingly distinct scenarios.

\subsection{Chiral vortex and Chiral vortex \\ $\mathrm{L}_{lk}(\mathcal{C}^1_{(1,1)},\mathcal{C}^2_{(1,1)})$ }

This process was explicitly related to topological Josephson junctions at the edge of quantum spin-Hall insulators. We have already provided a topological argument showing that the effect of a \emph{single} link ($lk=1$) between chiral vortices has as consequence a generation of a pair of zero modes, one in each string, due to a boundary condition change from periodic to anti-periodic. Proceeding further, our Josephson junction picture emphasizes that a double link ($lk=2$) between chiral vortices does not lead simply to another twist of boundary conditions and the loss of zero-modes obtained in the first link; rather, a second link between the loops is equivalent to a $2\pi$ evolution of the Josephson phase for a junction at the edge of a quantum spin-Hall insulator. Thus, the second linking leads to the exchange of a $e/2$ quasi-particle between the involved links. Generalizing, an odd number $2n+1$ of linking events between chiral vortices leads to the exchange of $ne/2$ charges and a zero mode, while an even number $2n$ exchanges singly $ne/2$ charges. Importantly, since the strings are immersed into a superconducting medium, a total of $2n=8$ links implies the exchange of a total of $2e$ charges, which can be absorbed in the superconducting condensate. This implies a topological redundancy in the linking processes of chiral vortices, which are then classified as $\mathbb{Z}_8$, that is, after 8 links, from the point of view of the internal structure of the strings, they are equivalent to unlinked loops.

As a final comment, it has been argued that vortices of type $(1,1)$ cannot mutually link in the context of time-reversal invariant topological superconductors, as they are restricted to live on such system's boundaries \cite{GuQi}. We refrain from such subtleties, not restricting ourselves to  time-reversal symmetric systems; we are concerned only with the general features of linking between chiral vortices. Also, as shown below, other distinct scenarios, which are not constrained by the discussions of Ref. [\onlinecite{GuQi}], are in fact topologically equivalent to this one, making it a valuable starting point of analysis, due to its simplicity.

\subsection{Dirac string and trivial vortex \\ $\mathrm{L}_{lk}(\mathcal{C}^1_{(2,0)},\mathcal{C}^2_{(0,2)})$}

As previously discussed, the double link of chiral vortices returns the same topological invariant as simple links between Dirac strings and trivial vortices, $\mathrm{L}_{2lk}(\mathcal{C}^1_{(1,1)},\mathcal{C}^2_{(1,1)})=\mathrm{L}_{lk}(\mathcal{C}^1_{(2,0)},\mathcal{C}^2_{(0,2)})$. Indeed, the latter process corresponds to the introduction of a $\pi$-flux into a pair of chiral Majorana modes, and in the Josephson-junction picture amounts to a $\pi$ evolution of the Josephson phase. Again, this process therefore involves the exchange of an $e/2$ charge between the trivial vortex and Dirac string.

\subsection{Chiral vortex and trivial vortex \\ $\mathrm{L}_{lk}(\mathcal{C}^1_{(1,1)},\mathcal{C}^2_{(0,2)})$}

A simple way to study this problem is to consider the additive properties of the link invariant as discussed in Sec. \ref{sec:linking}. Decomposing the trivial vortex in terms of chiral vortices of opposite chirality, 
\begin{equation}
\mathcal{C}^2_{(0,2)}=\mathcal{C}^2_{(1,1)} \cup \mathcal{C}^2_{(-1,1)},
\end{equation} the linking at matter is the same as discussed for a single linking between chiral vortices, if one notices that
\begin{equation}
\mathrm{L}_{lk}(\mathcal{C}^1_{(1,1)},\mathcal{C}^2_{(-1,1)})=0,
\end{equation}
i.e., the linking invariant between chiral vortices of opposite chirality vanishes. Following our present analysis, therefore, we conclude that linkings between chiral vortices and trivial vortices also reduce to our original scenario, and that 8-fold linkings between them also return a trivial process.

\section{Further fractionalization}\label{sec:fractionalize}

In this section, we discuss further fractionalization possibilities which can be facilitated by strong many-body interactions. The vortices we have been considering so far have restricted topological indices, namely $N_\Phi+N$ is even. This is a result from the fact that all vortices can be decomposed into combinations of the two primitive chiral vortices, which have $(N_\Phi,N)=(1,1)$ and $(1,-1)$ respectively. In other words, a chiral Majorana string (or in general a string with half-integral chiral central charge) must be accompanied by a half-flux quantum $\Phi=hc/4e$ (resp.~an odd multiple of $hc/4e$). Here we seek fractional vortices that allow $N_\Phi+N$ to be odd. In this instance, the smallest fractional vortices will have $(N_\Phi,N)=(1,0)$ and $(0,1)$. The former will only carry a magnetic flux $\Phi=hc/4e$ and the latter will be a pure chiral Majorana string. The fractionalization will therefore completely decouple or deconfine all electronic chiral modes from magnetic fluxes. We will conclude this section by speculating the fractional quasiparticle exchange through linking between such fractional vortices.

We begin with the superconducting Dirac semimetal Hamiltonian \eqref{Dirac-SC} with the pairing $\Delta\Gamma$ as then. It was then seen to be convenient to decompose the pairing parameters in $\Delta_\pm$, each corresponding to the pairing energy gap of a Weyl fermion with chirality $\tau_z=\pm1$, respectively. When the pairing phases $\varphi_\pm$ of the pairing parameters $\Delta_\pm=|\Delta_\pm|e^{i\varphi_\pm}$ are $0$ or $\pi$, the system is time reversal symmetric and falls into class DIII of the Altland-Zirnbauer classification\cite{AltlandZirnbauer97} of electronic band theories. Along a surface interface separating two domains with pairing phases $(\varphi_+,\varphi_-)=(0,0)$ and $(\pi,0)$, there lies a massless 2D Majorana fermion protected by time reversal. This is identical to the anomalous boundary surface state of a class DIII topological superconductor\cite{SchnyderRyuFurusakiLudwig08,Kitaevtable08,Ryu2010ten,QiHughesRaghuZhang09} in 3D such as the superfluid-B phase of He$^3$.

\begin{figure}[htbp]
\centering\includegraphics[width=0.4\textwidth]{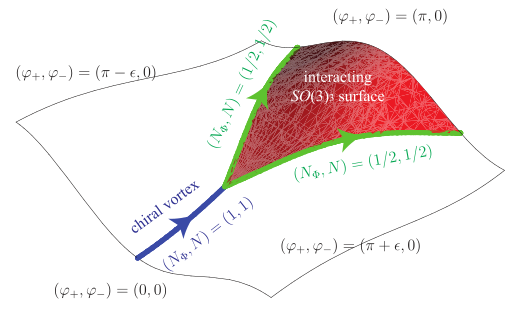}
\caption{Fractionalization of a chiral vortex on a surface interface. (Color online)}\label{fig:fracfig1}
\centering\includegraphics[width=0.5\textwidth]{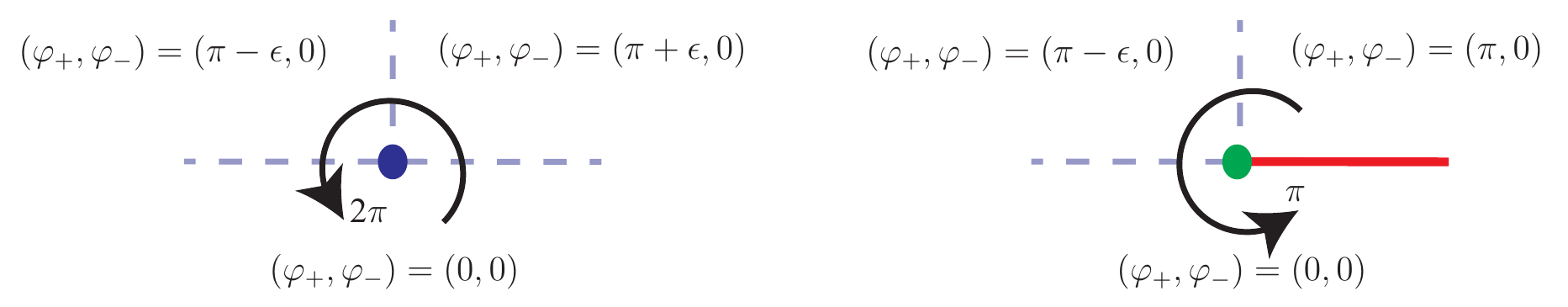}
\caption{Pairing parameters around a chiral vortex (left) and a fractional line interface (right). (Color online)}\label{fig:fracfig2}
\end{figure}

This Majorana surface state can acquire a time reversal breaking mass by having imaginary pairing; this mass can either be a positive or a negative number, and say we have this by shifting, either, $\delta\varphi_+=\pi\pm\epsilon$, for concreteness. Alternatively, the surface Majorana fermion can also gain a time reversal symmetric many-body mass gap through strong interactions.\cite{LukaszChenVishwanath} In this case the interacting gapped surface also carries a long-range entangled fermionic topological order and supports a $SO(3)_3$ (or $SU(2)_6$) like anyon structure, which also manifests as protected boundary degrees of freedom. Appendix \ref{app:SONWZW} contains some general information about $SO(N)_1$ Wess-Zumino-Witten CFTs, intended as a review of properties and relations used to understand our following results.

We may thus consider a tri-junction separating such three distinct massive surface states, as depicted in Figure~\ref{fig:fracfig1}. The line interface (blue line) between the two time reversal breaking surfaces with $\delta\varphi_+=\pi\pm\epsilon$ hosts a chiral Majorana mode. Moreover, the pairing phase $\varphi_+$ winds by $2\pi$ once around the line defect when it is embedded in 3D. Such surface domain wall therefore represents a chiral vortex with magnetic flux $\Phi=hc/4e$ and chiral central charge $c=1/2$, i.e.~$(N_\Phi,N)=(1,1)$.

The chiral vortex is then non-locally split and fractionalized into a pair of smaller components (green lines). They are facilitated by the insertion of the many-body interacting symmetric $SO(3)_3$ (red) surface. From the conservation of energy-momentum and magnetic field, each fractional line interface carries half of the magnetic flux and thermal current, i.e.~$(N_\Phi,N)=(1/2,1/2)$ (and chiral central charge $c=1/4$). For instance the pairing phase $\varphi_+$ winds only by $\pi$ once around a fractional line interface (see figure~\ref{fig:fracfig2}). 

\subsection{The $\mathcal{G}=SO(3)_3\boxtimes\overline{SO(4)_1}$ conformal field theory}\label{sec:GCFT}

The low energy modes running along the fractional vortex can be described by a conformal field theory with chiral central charge $c=1/4$. One possible theory is given by the relative tensor product $\mathcal{G}=SO(3)_3\boxtimes\overline{SO(4)_1}$.\cite{SahooZhangTeo15} Here we review the primary field (fundamental excitation) content of this theory. The $SO(3)_3$ (or equivalently $SU(2)_6$) theory has 7 primary fields labeled by $j={\bf 0},{\bf 1/2},\ldots,{\bf 3}$ with conformal dimensions $h_j=j(j+1)/8$. For instance, ${\bf 3}$ behaves like a Majorana fermion. The Monodromy phase between a primary field $j$ with the fermion ${\bf 3}$ gives 1 if $j$ is an integer, or $-1$ if $j$ is a half-integer. Thus in a closed ring geometry, the half-integral primary field excitations $j={\bf 1/2},{\bf 3/2},{\bf 5/2}$ must come in pairs, unless there exists another chiral vortex of the same Weyl fermion that links the fractional vortex. In this case, the chiral vortex changes the periodic boundary condition of the fractional vortex by an additional $-1$ phase, and forces odd multiples of the half-integral primary fields.

The $SO(3)_3$ theory however has a chiral central charge of $9/4$, which is off the appropriate chiral central charge $1/4$ by 2. This is compensated by the $\overline{SO(4)_1}$ sector, which is represented with the overline to indicate that it is assumed to propagate in the opposite direction. The $\overline{SO(4)_1}=\overline{SU(2)_1}\times\overline{SU(2)_1}$ theory is Abelian and has a bosonized description. $\mathcal{L}_{\overline{SO(4)_1}}=\frac{1}{2\pi}K_{IJ}\partial_x\phi^I\partial_t\phi^J-V_{IJ}\partial_x\phi^I\partial_x\phi^J$, where $I,J=1,2$ labels the two decoupled $\overline{SU(2)_1}$ sectors and $K_{IJ}=-2\delta_{IJ}$ is the Cartan matrix of $\overline{SO(4)_1}$. There are 4 Abelian primary fields $1,\bar{s}_1,\bar{s}_2,\bar{s}_1\bar{s}_2$, where $\bar{s}_I$ are semions with conformal dimension $h_s=-1/4$.

The semion pair $\bar{s}_1\bar{s}_2$ is fermionic and should be identified with $j={\bf 3}$ in the $SO(3)_3$ sector to represent the Majorana fermion. The identification is achieved by condensing\cite{BaisSlingerlandCondensation,SahooZhangTeo15} the Abelian boson ${\bf 3}\otimes\bar{s}_1\bar{s}_2$ in the tensor product theory $SO(3)_3\otimes\overline{SO(4)_1}$. The condensed theory, denoted by the relative tensor product $SO(3)\boxtimes\overline{SO(4)_1}$, now has the appropriate chiral central charge $c=9/4-2=1/4$. It has 7 primary fields,  and is $\mathbb{Z}_2$ graded into two parts $\mathcal{G}_0$ and $\mathcal{G}_1$. $\mathcal{G}_0$ contains the 4 integral primary fields $j={\bf 0},{\bf 1},{\bf 2},{\bf 3}$, which we will now respectively relabel by $1,\gamma_+,\gamma_-,f$, that are relatively local (i.e. have trivial monodromy) with the Majorana fermion $f\equiv{\bf 3}\equiv\bar{s}_1\bar{s}_2$. The condensate ${\bf 3}\otimes\bar{s}_1\bar{s}_2$ further identifies, for instance, $\gamma_+\equiv{\bf 1}\equiv{\bf 2}\otimes\bar{s}_1\bar{s}_2$ and $\gamma_-\equiv{\bf 2}\equiv{\bf 1}\otimes\bar{s}_1\bar{s}_2$. The two $\gamma_\pm$ excitations (not to be confused with the Majorana fields of our previous non-interacting discussions) are non-Abelian and obey the fusion rules \begin{gather}f\times f=1,\quad\gamma_\pm\times f=\gamma_\mp\nonumber\\\gamma_\pm\times\gamma_\pm=1+\gamma_++\gamma_-.\label{G0fusion}\end{gather}

The odd sector $\mathcal{G}_1$ contains the remaining 3 half-integral primary fields $\alpha_+\equiv{\bf 1/2}\otimes\bar{s}_1,\beta\equiv{\bf 3/2}\otimes\bar{s}_1,\alpha_-\equiv{\bf 5/2}\otimes\bar{s}_1$ that are non-local (i.e. have non-trivial, $-1$, monodromy) with the fermion $f$. The half-integral fields alone would not be local  with the boson condensate ${\bf 3}\otimes\bar{s}_1\bar{s}_2$ if they were not tensored with the semion $\bar{s}_1$. The boson condensate identifies $\alpha_+\equiv{\bf 5/2}\otimes\bar{s}_2,\beta\equiv{\bf 3/2}\otimes\bar{s}_2,\alpha_-\equiv{\bf 1/2}\otimes\bar{s}_2$. They obey the fusion rules \begin{gather}f\times\alpha_\pm=\alpha_\mp,\quad f\times\beta=\beta\nonumber\\\beta\times\gamma_\pm=\alpha_++\alpha_-+\beta\nonumber\\\beta\times\beta=1+\gamma_++\gamma_-+f\label{G1fusion}\\\alpha_\pm\times\gamma_\pm=\alpha_++\beta,\quad\alpha_\pm\times\beta=\gamma_++\gamma_-\nonumber\\\alpha_\pm\times\alpha_\pm=1+\gamma_+.\nonumber\end{gather} Their conformal dimensions and quantum dimensions are listed in table~\ref{tab:Gdim}. The modular $S$-matrix is given by \begin{align}S_{ij}=(-1)^{ij}\sin\left(\frac{\pi(i+1)(j+1)}{8}\right)\end{align} where $i,j=0,1,\ldots,6$ are ordered according to the 7 primary fields $1,\alpha_+,\gamma_+,\beta,\gamma_-,\alpha_-,f$. 
\begin{table}[htbp]
\centering
\begin{tabular}{l|lllllll}
${\bf x}$&1&$\alpha_+$&$\gamma_+$&$\beta$&$\gamma_-$&$\alpha_-$&$f$\\\hline
$d_{\bf x}$&1&$\sqrt{2+\sqrt{2}}$&$1+\sqrt{2}$&$\sqrt{4+2\sqrt{2}}$&$1+\sqrt{2}$&$\sqrt{2+\sqrt{2}}$&1\\
$h_{\bf x}$&0&$\frac{27}{32}$&$\frac{1}{4}$&$\frac{7}{32}$&$\frac{3}{4}$&$\frac{27}{32}$&$\frac{1}{2}$\\
$\mathcal{M}_{{\bf x}f}$&1&$-1$&1&$-1$&1&$-1$&1
\end{tabular}
\caption{The conformal dimension $h_{\bf x}$ (mod 1), quantum dimension $d_{\bf x}$ and monodromy phase $\mathcal{M}_{{\bf x}f}$ with the fermion of primary fields ${\bf x}$ in the $\mathcal{G}=SO(3)_3\boxtimes\overline{SO(4)_1}$ CFT.}\label{tab:Gdim}
\end{table}

We notice that the $\mathcal{G}$ CFT has a $\mathbb{Z}_2$ grading structure $\mathcal{G}=\mathcal{G}_0\oplus\mathcal{G}_1$ where primary fields ${\bf x}$ are separated by the monodromy phase $\mathcal{M}_{{\bf x}f}=\pm1$ with the fermion $f$ (see table~\ref{tab:Gdim}). The grading is compatible with fusion in the sense that $\mathcal{G}_\sigma\times\mathcal{G}_\tau\to\mathcal{G}_{\sigma+\tau}$ for $\sigma,\tau=0,1$ mod 2. Moreover the two sectors have identical quantum dimensions \begin{align}\mathcal{D}_{\mathcal{G}_\sigma}=\sqrt{\sum_{{\bf x}\in\mathcal{G}_\sigma}d_{\bf x}^2}=2\sqrt{4+2\sqrt{2}}.\end{align} In a closed ring geometry with a fixed fermion boundary condition, the CFT is restricted to one of the two sectors.

\subsection{Fractional vortices and fractional linking}
The fractional vortex we discussed in the previous subsection involves a $\pi$ winding of pairing phase $\varphi_+$ of the positive chiral Weyl fermion. It has topological index $(N_\Phi,N)=(1/2,1/2)$. Similarly one can introduce a many-body interacting gap at a $\pi$-junction for the negative chiral Weyl fermion. Such a $\pi$ pairing vortex of the negative chiral Weyl fermion has topological index $(N_\Phi,N)=(1/2,-1/2)$, where the low energy modes propagate in the opposite direction of the magnetic flux. This can be shown by reversing the propagating directions as well as replacing $\varphi_+\leftrightarrow\varphi_-$ in figure~\ref{fig:fracfig1}. Combining these fracional vortices with opposite chiralities, we arrive at the two primitive fractional vortices hinted in the introduction of this section: \begin{align}\mbox{Half-vortex: }(N_\Phi,N)&=(1,0)\nonumber\\&=(1/2,1/2)+(1/2,-1/2)\nonumber\\\mbox{Majorana string: }(N_\Phi,N)&=(0,1)\nonumber\\&=(1/2,1/2)+(-1/2,1/2)\nonumber\end{align}

The half-vortex carries counter-propagating low energy CFT's $\mathcal{G}\otimes\overline{\mathcal{G}}$ which are not protected. Noticing the current algebra structure in $\mathcal{G}=SO(3)_3\boxtimes\overline{SO(4)_1}$ and $\overline{\mathcal{G}}=\overline{SO(3)_3}\boxtimes SO(4)_1$, the low energy modes can gain a many-body mass by the backscattering interaction\cite{SahooZhangTeo15} \begin{align}V=u\left({\bf J}_{SO(3)_3}\cdot{\bf J}_{\overline{SO(3)_3}}+{\bf J}_{SO(4)_1}\cdot{\bf J}_{\overline{SO(4)_1}}\right)\end{align} where ${\bf J}=(J^1,\ldots,J^D)$ are the Kac-Moody current operators of the Affine Lie algebras, for $D=3$ or 6 the dimensions of $SO(3)$ or $SO(4)$.

On the other hand, the magnetic flux cancels along the the Majorana string whereas the two low energy CFT's are co-propagating. We denote the combination of the the CFT's by the relative tensor product $\mathcal{F}=\mathcal{G}^I\boxtimes_b\mathcal{G}^{II}$. Each sector $a=I,II$ consists of the even primary fields $\mathcal{G}^a_0=\langle1,\gamma_+^a,\gamma_-^a,f^a\rangle$ and the odd primary fields $\mathcal{G}^a_1=\langle\alpha_+^a,\alpha_-^a,\beta^a\rangle$ with respect to the fermion $f^a$ (see subsection~\ref{sec:GCFT}). The two fermions $f^I$ and $f^{II}$ should be identified with  $f$, the single Majorana fermion that should remain. The label $b$, moreover, indicates that the tensor product contains the mutually local bosonic pairs \begin{align}b=\left\{f^I\otimes f^{II},\gamma_+^I\otimes\gamma_-^{II},\gamma_-^I\otimes\gamma_+^{II}\right\}.\label{bcondensate}\end{align} The relative tensor product $\mathcal{G}^I\boxtimes_b\mathcal{G}^{II}$ is a result from condensing the above bosonic set. The fusion content of the condensed theory was presented in Ref.~\onlinecite{SahooZhangTeo15}. Here we review the relevant features of the condensation following the procedure proposed in Ref.~\onlinecite{BaisSlingerlandCondensation}.

First the condensation of the fermion pair $f^If^{II}$ identifies the fermions and the semions \begin{align}f\equiv f^I\equiv f^{II},\quad g_\pm\equiv\gamma_\pm^I\equiv\gamma_\mp^{II}.\end{align} The condensation of the non-Abelian boson $\gamma_+^I\gamma_-^{II}$ 
results in the fusion rule 
\begin{align}
g_\pm\times g_\pm=1+g_++g_-,\quad g_\pm\times f=g_\mp.
\end{align} 
Under these identifications, the tensor product $\mathcal{G}^I\otimes\mathcal{G}^{II}$ reduces to the $\mathbb{Z}_2\times\mathbb{Z}_2$-graded fusion theory \begin{align}\mathcal{F}'=\mathcal{F}'_{00}\oplus\mathcal{F}'_{11}\oplus\mathcal{F}'_{10}\oplus\mathcal{F}'_{01},\label{Fgrading}\end{align} where the components are generated by $\mathcal{F}'_{00}=\langle1,f,g_\pm\rangle$, $\mathcal{F}'_{11}=\langle\sigma,a\rangle$, $\mathcal{F}'_{10}=\langle\alpha_\pm^I,\beta^I\rangle$ and $\mathcal{F}'_{01}=\langle\alpha_\pm^{II},\beta^{II}\rangle$. Here $\sigma$ and $a$ descend from the original tensor product theory by identifying \begin{align}
a&\equiv\alpha_\pm^I\alpha_\pm^{II}\equiv\alpha_\pm^I\alpha_\mp^{II}\nonumber\\\alpha_\pm^I\beta^{II}&\equiv\beta^I\alpha_\pm^{II}\equiv\sigma+a.\label{absigma}\end{align} 
For instance, the new field $\sigma$ behaves exactly like the Ising twist field with conformal dimension $h_\sigma=h_\alpha+h_\beta=27/32+7/32\equiv1/16$ (mod 1). It also obeys the appropriate fusion rule \begin{align}\sigma\times\sigma=1+f,\quad\sigma\times f=\sigma.\end{align} Moreover, it can be shown that $a$ can be generated by combining \begin{align}\sigma\times g_\pm=a.\end{align} To complete the fusion theory, other fusion rules can be deduced from the original ones in $\mathcal{G}_I\otimes\mathcal{G}_{II}$ together with the modified $\beta^I\beta^{II}=2a$.

The $\mathbb{Z}_2\times\mathbb{Z}_2$ grading structure \eqref{Fgrading} is defined according to the monodromy phases $\mathcal{M}_{{\bf x}f^I},\mathcal{M}_{{\bf x}f^{II}}=\pm1$ with the two fermions $f^I$ and $f^{II}$. Each component carries the identical quantum dimension \begin{align}\mathcal{D}_{\mathcal{F}'_{\sigma\tau}}=\sqrt{\sum_{{\bf x}\in\mathcal{F}'_{\sigma\tau}}d_{\bf x}^2}=2\sqrt{2+\sqrt{2}}.\end{align} Among the fusion theory $\mathcal{F}'$, all fields are confined by the condensate \eqref{bcondensate}, except for \begin{gather}\mathcal{F}=\mathcal{G}^I\boxtimes_b\mathcal{G}^{II}=\mathcal{F}_0\oplus\mathcal{F}_1\nonumber\\\mathcal{F}_0=\langle1,f\rangle,\quad\mathcal{F}_1=\langle\sigma\rangle,\end{gather} which is identical to the primary field content of the chiral Ising CFT (or chiral Majorana fermion). This justifies the terminology of calling the pair $\mathcal{G}^I\boxtimes_b\mathcal{G}^{II}$ a Majorana string. 

The other confined fields in $\mathcal{F}'$ in \eqref{Fgrading} are not primary field excitations of the Ising CFT. They should be treated as extrinsic defects. In a closed ring geometry, the antiperiodic fermionic boundary condition only allows the primary fields in $\mathcal{F}_0=\langle1,f\rangle$ to appear. When linking the Majorana string $\mathcal{C}^1$: $(N_\Phi,N)=(0,1)$ with a regular non-chiral vortex $\mathcal{C}^2$: $(N_\Phi,N)=(2,0)$, the boundary condition changes to periodic, enforcing the presence of a zero mode and the twist field sector $\mathcal{F}_1=\langle\sigma\rangle$. The exchange of a Majorana zero mode agrees with the expectation from the linking invariant \eqref{lkinv1} $L(\mathcal{C}^1,\mathcal{C}^2)=1$.

When linking the Majorana string $\mathcal{C}^1$: $(N_\Phi,N)=(0,1)$ with a chiral vortex $\mathcal{C}^2$: $(N_\Phi,N)=(1,1)$, only the 1st fermion $f^I$ has a change in the boundary condition. This fractional boundary condition twists the Ising CFT to the defect sector $\mathcal{F}'_{10}=\langle\alpha_\pm^I,\beta^I\rangle$ in \eqref{Fgrading}. From \eqref{absigma}, we see that the non-Abelian $\alpha$ and $\beta$ fields are fractionalizations of the Ising twist field $\sigma$, and therefore in a sense, half a zero mode is exchanged in the linking process. This matches with the linking invariant $L(\mathcal{C}^1,\mathcal{C}^2)=1/2$.



\section{Conclusion and discussion}\label{sec:conc}

The classification of topological phases in the presence of interactions remains an open problem. Just as in the non-interacting limit\cite{TeoKaneDefects}, we see that the introduction of spatial defects introduces a new layer of subtleties to this problem also in the interacting case. The defects carry internal structure and the rearrangement of this structure under manipulations of the defects allow for a topological classification in terms of their protected modes and quasi-particle exchanges. As we showed, such classification is sensitive to the effects of interactions. 

We considered one-dimensional defects, strings, immersed in s-wave superconducting Dirac metals. We showed existence of three distinct types of vortices, namely regular vortices, Dirac strings and chiral vortices, which can be used as building blocks to any more complicate scenario. They are classified in terms of a pair of numbers $(N_\Phi,N)$, which count the amount of $hc/4e$ and of chiral Majorana modes, respectively, carried by a given vortex. From these quantities, we introduced a topological invariant, called the linking invariant, characterized by a counting of the distinct number of links between vortex loops, as well as their inners structures. With our invariant, a topological classification of vortex strings under linking processes was provided.

The linking of closed loops is a non-smooth process which connects two topologically distinct situations. Previously described by means of instanton calculations and topological electromagnetic responses\cite{QWZ,GuQi,StoneLopes}, their description in the superconducting scenarios becomes more complicated, owing to the existence of electromagnetically neutral chiral Majorana modes propagating along the vortices (we emphasize the puzzling nature of the problem due to the charge neutral Majorana mode). Our picture of describing these topologically non-smooth phenomena in terms of Josphson effects offers a new alternative to approach the problem, which as we showed allows even for an explicit and microscopic consideration of effects of interactions. In particular, it allowed us showing that the vortices in these systems follow a $\mathbb{Z}_8$ classification. Several connections are then hinted by our work, relating topological classifications, pumps, defects, Josephson effects, instanton processes; the particular relation between Josephson effects and instanton processes is particularly remarkable.

Our extension of the problem to strong interactions also adds to the zoo of exotic fractional quasi-particles. While we cannot prove that our anyon condensation picture is unique, it is based on the Bais-Slingerland condensation procedure\cite{BaisSlingerlandCondensation,SahooZhangTeo15}, and is both the simplest possibility and is consistent with all of our structures. In particular, its consistency with our defined topological invariant is a strong suggestion that our process is following the correct direction.

Questions remains that can be approached. Whether the relationship between Josephson effects and topological instanton processes can be extended to other scenarios is one of them. Also remains the possibility of relating the classification of the distinct types of strings, which is a 1D defect classification, to the $\mathbb{Z}_{16}$ classification of bulk time-reversal topological superconductors, which is a bulk 3D classification related by bulk-boundary correspondence to surfaces as 2D defects. Regarding this last point, the consideration of quasi-particles spins, not considered here yet, may be of importance.

\section*{Acknowledgements}

We would like to thank Michael Stone for insightful comments and conversations. PLSL is supported by the Canada First Research Excellence Fund. This work is supported by the NSF under Grant No. DMR-1455296 (SR) and DMR-1653535 (JCYT).

\appendix
\section{Chiral modes in SC Weyl fermions} \label{app:appendix}

The main characters in our analysis of vortex-linking problems are the carried magnetic fluxes the inner structure of chiral bound states in vortices of s-wave superconducting Weyl Hamiltonians. Here we discuss in detail the physics of vortex bound states in this context, precisely showing that the lowest energy ones correspond to electrically neutral Majorana chiral modes  \eqref{eq:WeylSCH}. We start with a review of known results and then extend them both analytically and numerically.

\subsection*{Algebraic radial BdG Hamiltonian}

Let us display a general procedure which allows the analytical, in some limits, and numerical, generally, evaluation of the mean-field superconducting states in the presence of a vortex. We follow the general lines of \cite{PedroPouyan,Gygi}. 

Start with a superconducting Weyl fermion in 3D with a vortex along the z-direction. The full second quantized Hamiltonian,
in the presence of a vortex and a background EM field, reads
\begin{equation}
H=\int d^{3}x\Psi^{\dagger}h_{BdG}\Psi,
\end{equation}
where
\begin{eqnarray}
h_{BdG} & = & v_{F}\boldsymbol{\Gamma}\cdot\left(-i\nabla\right)+\Lambda_{x}\Delta_{0}\left(r\right)e^{i\rho_{z}\theta}
\nonumber \\
 &  & -v_{F}\rho_{z}\boldsymbol{\Gamma}\cdot\mathbf{A}+\rho_{z}A_{0}
 \nonumber \\
 & = & v_{F}\boldsymbol{\Gamma}\cdot\left(-i\nabla\right)_{\perp}+\Lambda_{x}\Delta_{0}\left(r\right)e^{i\rho_{z}\theta}
 \nonumber \\
 &  & +v_{F}\Gamma_{z}\left(-i\partial_{z}\right)
 \nonumber \\
 &  & -v_{F}\rho_{z}\boldsymbol{\Gamma}\cdot\mathbf{A}+\rho_{z}A_{0}
 \nonumber \\
 & \equiv & h_{\perp}+h_{z}+h_{EM}.
\end{eqnarray}
Our basis has $\boldsymbol{\Gamma}=\rho_z \boldsymbol{\sigma}$ and $\Lambda_x=\rho_x$, with $\rho$ and $\sigma$ Pauli matrices in the Nambu and spin spaces, respectively. It will prove enough to set the $k_{z}$-dispersion to zero and treat it perturbatively in each bound mode, in order to see the chiral structure appearing. We also start at the limit of vanishing external gauge fields, returning to them later, in order to study the electromagnetic coupling and charge of the vortex bound state.

So we take $k_z =\boldsymbol{A}=A_0=0$ and make use of the assume rotational symmetry along the z-direction, writing wavefunctions as
\begin{equation}
\psi_{l}\left(r,\theta\right)=e^{i\left(l-\frac{\rho_{z}+\sigma_{z}}{2}\right)\theta}\phi\left(r\right).
\end{equation}
The function $\phi\left(r\right)$ satisfies the radial superconducting Weyl equation at $k_z=0$, obtained from $h_{\perp}$, which reads
\begin{widetext}
\begin{equation}
\left(\begin{array}{cccc}
-\bar{\mu} & -i\left[\partial_{\bar{r}}+\frac{l}{\bar{r}}\right] & \left|\bar{\Delta}\right|\\
-i\left[\partial_{\bar{r}}-\frac{\left(l-1\right)}{\bar{r}}\right] & -\bar{\mu} &  & \left|\bar{\Delta}\right|\\
\left|\bar{\Delta}\right| &  & \bar{\mu} & i\left[\partial_{\bar{r}}+\frac{\left(l+1\right)}{\bar{r}}\right]\\
 & \left|\bar{\Delta}\right| & i\left[\partial_{\bar{r}}-\frac{l}{\bar{r}}\right] & \bar{\mu}
\end{array}\right)\phi_{nl}\left(\bar{r}\right)=E_{nl}\phi_{nl}\left(\bar{r}\right),
\end{equation}
\end{widetext}
where we rescaled the energies by the gap strength $\Delta_{0}$ and
lengths by the coherence length $v_{F}/\Delta_{0}$; then, labeling
the dimensionless variables by bars. In what follows we omit the bars to avoid cluttering.

Define the kinetic Hamiltonian
\begin{equation}
K_{l}\equiv\left(\begin{array}{cc}
-\mu & -i\left[\partial_{r}+\frac{l}{r}\right]\\
-i\left[\partial_{r}-\frac{\left(l-1\right)}{r}\right] & -\mu
\end{array}\right).
\end{equation}
We expand the wavefunctions in terms of Bessel functions. Noticing
the following recurrence relations
\begin{eqnarray}
\left(\partial_{r}+\frac{l}{r}\right)J_{l}\left(kr\right) & = & kJ_{l-1}\left(kr\right),\\
\left(\partial_{r}-\frac{l}{r}\right)J_{l}\left(kr\right) & = & -kJ_{l+1}\left(kr\right),
\end{eqnarray}
we may define the raising and lowering operators
\begin{eqnarray}
a_{l} & = & \left(\partial_{r}+\frac{l}{r}\right)\\
a_{l}^{\dagger} & = & -\left(\partial_{r}-\frac{l-1}{r}\right).
\end{eqnarray}
They naturally act as
\begin{eqnarray}
a_{l}J_{l}\left(kr\right) & = & kJ_{l-1}\left(kr\right)\\
a_{l}^{\dagger}J_{l-1}\left(kr\right) & = & kJ_{l}\left(kr\right)
\end{eqnarray}
and allows writing the Bessel equation naturally as
\begin{eqnarray}
a_{l}^{\dagger}a_{l}J_{l}\left(kr\right) & = & k^{2}J_{l}\left(kr\right).
\end{eqnarray}
In this language, the kinetic Hamiltonian reads
\begin{equation}
K_{l}\equiv\left(\begin{array}{cc}
-\mu & -ia_{l}\\
ia_{l}^{\dagger} & -\mu
\end{array}\right).
\end{equation}

This suggests that eigenstates of the kinetic Hamiltonian may be written as $\left(\alpha J_{l-1}(kr),\beta J_l (kr)\right)^T$ which can be computed by solving
\begin{equation}
\left(\begin{array}{cc}
-\mu-\lambda & -ik\\
ik & -\mu-\lambda
\end{array}\right)\left(\begin{array}{c}
\alpha\\
\beta
\end{array}\right)=0
\end{equation}
Solutions give energies $\lambda_{l}^{\pm}\left(k\right)=-\mu\pm k$.
The corresponding coefficients are
\begin{equation}
\left(\begin{array}{c}
\alpha^{\pm}\\
\beta^{\pm}
\end{array}\right)=\frac{1}{\sqrt{2}}\left(\begin{array}{c}
1\\
\pm i
\end{array}\right).
\end{equation}
We thus write
\begin{equation}
\chi_{kl}^{\pm}=\frac{1}{\sqrt{\mathcal{N}_{k}}}\left(\begin{array}{c}
J_{l-1}\left(kr\right)\\
\pm iJ_{l}\left(kr\right)
\end{array}\right).
\end{equation}
The normalization factor reads
\begin{equation}
\mathcal{N}_{k}=\int_{0}^{\infty}rdrJ_{l}\left(kr\right)J_{l}\left(kr\right).
\end{equation}
In this continuum limit,
\begin{eqnarray}
 &  & \int_{0}^{\infty}rdrJ_{l}\left(kr\right)J_{l}\left(k'r\right)
 \nonumber \\
 & = & \int_{0}^{\infty}rdrJ_{l+1}\left(kr\right)J_{l+1}\left(k'r\right)
 \nonumber \\
 & = & \frac{\delta(k-k')}{k}
\end{eqnarray}
but the orthonormalization factors will be regularized in the case of a finite cylinder, which 
will be used for numerical computation of the solutions.

Nambu partner wavefunctions are derived similarly by diagonalizing $-K_{l+1}$.
The states read
\begin{equation}
\tilde{\chi}_{kl}^{\pm}=\frac{1}{\sqrt{\tilde{\mathcal{N}}_{k}^{\pm}}}\left(\begin{array}{c}
J_{l}\left(kr\right)\\
\pm iJ_{l+1}\left(kr\right)
\end{array}\right)
\end{equation}
with energies $\tilde{\lambda}_{l}^{\pm}\left(k\right)=\mu\pm k$.
Notice $\tilde{\lambda}_{l}^{\pm}=-\lambda_{l}^{\pm}$. We take the
same label $l$ in the eigenvalues $\lambda$ for both Nambu sectors
as we will sum over the same values of $k$ (even in the finite radius cylinder.) 

The eigenstates of the vortex BdG Hamiltonian thus read
\begin{equation}
\psi_{nl}\left(r,\theta\right)=e^{i\left(l-\frac{\rho_{z}+\sigma_{z}}{2}\right)\theta}\phi_{nl}\left(r\right),
\end{equation}
where
\begin{equation}
\phi_{nl}\left(r\right)=\left(\begin{array}{c}
\int dk\left[c_{lk}^{n+}\chi_{kl}^{+}+c_{lk}^{n-}\chi_{kl}^{-}\right]\\
\int dk\left[d_{lk}^{n+}\tilde{\chi}_{kl}^{+}+d_{lk}^{n-}\tilde{\chi}_{kl}^{-}\right]
\end{array}\right).
\end{equation}
This procedure allows writing the BdG equation in a purely algebraic form,
\begin{eqnarray}
 &  & \left(\begin{array}{cc}
K_{l} & \left|\Delta\right|\\
\left|\Delta\right| & -K_{l+1}
\end{array}\right)\left(\begin{array}{c}
\int_{k}\left[c_{lk}^{n+}\chi_{kl}^{+}+c_{lk}^{n-}\chi_{kl}^{-}\right]
\nonumber\\
\int_{k}\left[d_{lk}^{n+}\tilde{\chi}_{kl}^{+}+d_{lk}^{n-}\tilde{\chi}_{kl}^{-}\right]
\end{array}\right)\\
 & = & E_{nl}\left(\begin{array}{c}
\int_{k}\left[c_{lk}^{n+}\chi_{kl}^{+}+c_{lk}^{n-}\chi_{kl}^{-}\right]\\
\int_{k}\left[d_{lk}^{n+}\tilde{\chi}_{kl}^{+}+d_{lk}^{n-}\tilde{\chi}_{kl}^{-}\right]
\end{array}\right).
\end{eqnarray}
Taking the scalar product with $\phi_{nl}$ with a $\int rdr$ measure, and using
\begin{widetext}
\begin{eqnarray}
\int_{0}^{\infty}rdr\tilde{\chi}_{k'l}^{\sigma' \dagger}\tilde{\chi}_{kl}^{\sigma} & = & \delta_{\sigma\sigma'}\delta\left(k-k'\right)\\
\int_{0}^{\infty}rdr\chi_{k'l}^{\sigma' \dagger}\left|\Delta\right|\tilde{\chi}_{kl}^{\sigma} & = & \frac{1}{\sqrt{\mathcal{N}^{\sigma}\tilde{\mathcal{N}}^{\sigma'}}}\int_{0}^{\infty}rdr\left|\Delta\right|\left[J_{l-1}\left(k'r\right)J_{l}\left(kr\right)+\sigma\sigma'J_{l}\left(k'r\right)J_{l+1}\left(kr\right)\right]\nonumber \equiv\left|\Delta\right|_{l,kk'}^{\sigma\sigma'},
\end{eqnarray}
we end at the problem 
\begin{equation}
\int_{k,k'}\left(\begin{array}{c}
c_{lk}^{n+}\\
c_{lk}^{n-}\\
d_{lk}^{n+}\\
d_{lk}^{n-}
\end{array}\right)^{\dagger}\left(\begin{array}{cccc}
\left(\lambda_{lk}^{+}-E_{nl}\right)\delta_{kk'} &  & \left|\Delta\right|_{l,kk'}^{++} & \left|\Delta\right|_{l,kk'}^{+-}\\
 & \left(\lambda_{lk}^{-}-E_{nl}\right)\delta_{kk'} & \left|\Delta\right|_{l,kk'}^{-+} & \left|\Delta\right|_{l,kk'}^{--}\\
\left|\Delta\right|_{l,k'k}^{++} & \left|\Delta\right|_{l,k'k}^{-+} & \left(-\lambda_{lk}^{+}-E_{nl}\right)\delta_{kk'}\\
\left|\Delta\right|_{l,k'k}^{+-} & \left|\Delta\right|_{l,k'k}^{--} &  & \left(-\lambda_{lk}^{-}-E_{nl}\right)\delta_{kk'}
\end{array}\right)\left(\begin{array}{c}
c_{lk'}^{n+}\\
c_{lk'}^{n-}\\
d_{lk'}^{n+}\\
d_{lk'}^{n-}
\end{array}\right)=0.
\end{equation}
\end{widetext}

\begin{figure}
\includegraphics[width=0.7\linewidth]{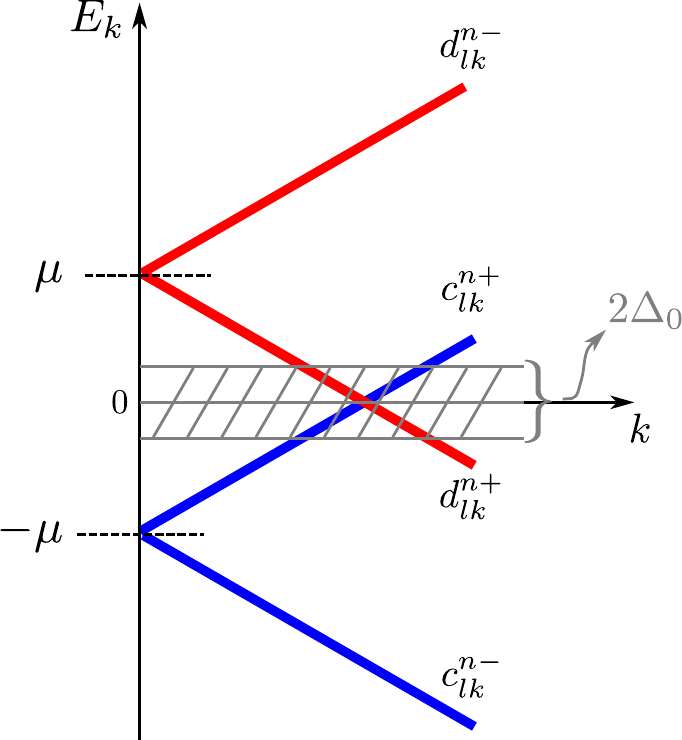}
\caption{Schematics of the Kinetic energy bands and coupling by superconductivity. Only the
bands $\lambda_{lk}^{+}$ and $-\lambda_{l+1k}^{+}$ overlap in energy
around the Fermi momentum. To compute the low-energy superconducting wavefunctions and vortex bound-states
spectrum numerically, it is enough to consider only these. In a finite system with radius
$R$, $k\rightarrow\alpha_{lj}/R$, where $\alpha_{lj}$ is the $j$-th zero of an order $l$ Bessel function. We focus in values of $j$ around
the lowest energy-bands crossing, grey region with the blue and red lines crossing. \label{fig:BandsRad}}
\end{figure}

Take a positive value for the chemical potential $\mu$. We are interested in the 
eigenstates of the problem induced by the superconducting pairing. The states
with kinetic energies $\lambda^{-}$ above are then high-energy and do not cross the
Fermi level for any value of $k$. They are not relevant for the
low energy excitations or superconducting pairing in general. One may then neglect the 
unnecessary coefficients, 
\begin{equation}
\phi_{nl}\left(r\right)\approx \left(\begin{array}{c}
\int dk\left[c_{lk}^{n+}\chi_{kl}^{+}\right]\\
\int dk\left[d_{lk}^{n+}\tilde{\chi}_{kl}^{+}\right]
\end{array}\right),
\end{equation}
in which case the eigenvalue problem simplifies to
\begin{widetext}
\begin{equation}
\int_{k,k^{\prime}}\left(\begin{array}{cc}
c_{lk}^{n+} & d_{lk}^{n-}\end{array}\right)\left(\begin{array}{cc}
\left(\lambda_{lk}^{+}-E_{nl}\right)\delta_{kk^{\prime}} & \left|\Delta\right|_{l,kk^{\prime}}^{++}\\
\left|\Delta\right|_{l,k^{\prime}k}^{++} & \left(-\lambda_{lk}^{+}-E_{nl}\right)\delta_{kk^{\prime}}
\end{array}\right)\left(\begin{array}{c}
c_{lk^{\prime}}^{n+}\\
d_{lk^{\prime}}^{n+}
\end{array}\right)=0.
\end{equation}
\end{widetext}

Quite generally, the complete set of states solving these consists of states with $E_{nl}\sim\Delta_{0}$, which
correspond to scattering states unbound to the vortex, and $E_{n_{CdG}l}=\Delta_{0}^{2}/E_{F}\times l$.
The wavefunctions read
\begin{eqnarray}
\psi_{nl}\left(r,\theta\right) & = & 
\int dk\left(\begin{array}{c}
\frac{c_{lk}^{n}}{\sqrt{\mathcal{N}_{k}}}\left(\begin{array}{c}
e^{i\left(l-1\right)\theta}J_{l-1}\left(kr\right)\\
ie^{il\theta}J_{l}\left(kr\right)
\end{array}\right)\\
\frac{d_{lk}^{n}}{\sqrt{\tilde{\mathcal{N}}_{k}}}\left(\begin{array}{c}
e^{il\theta}J_{l}\left(kr\right)\\
ie^{i\left(l+1\right)\theta}J_{l+1}\left(kr\right)
\end{array}\right)
\end{array}\right) \nonumber\\
 & \equiv & \left(\begin{array}{c}
\mathbf{u}_{ln}\left(r,\theta\right)\\
\mathbf{v}_{ln}\left(r,\theta\right)
\end{array}\right).
\end{eqnarray}
The normalization obeys $\mathcal{N}_{k}^{+}=\tilde{\mathcal{N}}_{k}^{-}\equiv\mathcal{N}_{k}$
and that we may discard the $+$ labels.

\subsection*{Finite size cylinder and numerics}

We may apply the previous results to a finite sized cylinder in order to implement a numerical diagonalization as a check that our diagonalization scheme works.

We set a finite size in the cylinder radius $R$. The wavefunction expansion then reads 
\begin{equation}
\phi_{nl}\left(r\right)=\left(\begin{array}{c}
\sum_{j}\left[c_{jl}^{n+}\chi_{jl}^{+}\right]\\
\sum_{j}\left[d_{jl}^{n+}\tilde{\chi}_{jl}^{+}\right]
\end{array}\right)
\end{equation}
with
\begin{align}
\chi_{lj}^{\pm}&=\frac{1}{\sqrt{\mathcal{N}_{lj}}}\left(\begin{array}{c}
J_{l-1}\left(\frac{\alpha_{lj}r}{R}\right)\\
\pm iJ_{l}\left(\frac{\alpha_{lj}r}{R}\right)
\end{array}\right),
\nonumber \\
\tilde{\chi}_{lj}^{\pm}&=\frac{1}{\sqrt{\mathcal{N}_{lj}}}\left(\begin{array}{c}
J_{l}\left(\frac{\alpha_{lj}r}{R}\right)\\
\pm iJ_{l+1}\left(\frac{\alpha_{lj}r}{R}\right)
\end{array}\right).
\end{align}
The momentum variables are exchanged as $k\to \alpha_{lj}$, where $\alpha_{lj}l$ is the $j$-th zero of the $l$-order Bessel function.
Making use of the Bessel functions identities, normalization reads,
\begin{eqnarray}
\mathcal{N}_{lj} & = & \left|RJ_{l+1}\left(\alpha_{lj}\right)\right|^{2},
\end{eqnarray} 
independently of the Nambu sector.

According to Fig.\ \ref{fig:BandsRad}, the index $j$ is an integer summed in a range $N_0$ (even for simplicity), from $-N_{0}/2$ up to
$N_{0}/2$ fixed at a starting value $N_{i}$ which fixes $\alpha_{lN_{i}}/R\sim k_{F}$.

The matrix elements now read approximately 
\begin{eqnarray}
 & & \int_{0}^{R}rdr\chi_{j'l}^{+\dagger}\chi_{jl}^{+}=\int_{0}^{R}rdr\tilde{\chi}_{j'l}^{+\dagger}\tilde{\chi}_{jl}^{+}=\delta_{jj'}\\
& &  \int_{0}^{R}rdr\chi_{k'l}^{\sigma'\dagger}\left|\Delta\right|\tilde{\chi}_{kl}^{\sigma}\nonumber \\ & = & \frac{\int_{0}^{1}xdx\left|\Delta\left(xR\right)\right|}{\sqrt{J_{l+1}\left(\alpha_{lj}\right)^{2}J_{l+1}\left(\alpha_{lj'}\right)^{2}}}\nonumber \\
 &  & \times \left[J_{l-1}\left(\alpha_{lj'}x\right)J_{l}\left(\alpha_{lj}x\right)+\sigma\sigma'J_{l}\left(\alpha_{lj'}r\right)J_{l+1}\left(\alpha_{lj}r\right)\right]\nonumber \\
 &  & \equiv\left|\Delta\right|_{l,jj'}^{\sigma\sigma'}.
\end{eqnarray}
and 
\begin{align}
&
\quad 
\left|\Delta\left(r\right)\right|  =  \Delta_{0}\tanh\left(\frac{r}{\xi}\right)
\nonumber \\
&\Rightarrow\left|\Delta\left(xR\right)\right|  =  \Delta_{0}\tanh\left(x\frac{R}{\xi}\right)
\end{align}
where $\xi=v_{F}\pi/\Delta_{0}$ is the SC coherence length and $\Delta_{0}$
is the magnitude of the SC pairing.

Thus, defining 
\begin{equation}
\Phi_{ln}=\left(c_{1l}^{n+}...c_{N_{0}l}^{n+},d_{1l}^{n+}...d_{N_{0}l}^{n+}\right),
\end{equation}
where 
$
\Phi_{ln}^{\dagger}\Phi_{ln'}=\delta_{nn'},
$
the eigenvalue problem becomes completely algebraic
\begin{eqnarray}
\left(\begin{array}{cc}
T_{l} & \Delta_{l}\\
\Delta_{l}^{T} & -T_{l}
\end{array}\right)\Phi_{ln} & = & E_{l,n}\Phi_{ln}
\end{eqnarray}
where
\begin{equation}
T_{l}=\left(\begin{array}{ccc}
\lambda_{l1}^{+}\\
 & \ddots\\
 &  & \lambda_{lN_{0}}^{+}
\end{array}\right)
\end{equation}
and
\begin{equation}
\Delta_{l}=\left(\begin{array}{ccc}
\left|\Delta\right|_{l,11}^{++} &  & \left|\Delta\right|_{l,1N_{0}}^{++}\\
 & \ddots\\
\left|\Delta\right|_{l,N_{0}1}^{++} &  & \left|\Delta\right|_{l,N_{0}N_{0}}^{++}
\end{array}\right).
\end{equation}

To solve for the eigenstates, all one needs to compute are the following integrals,
\begin{align}
I_{jj'}^{1} & =  \int_{0}^{1}xdx\left|\Delta\left(xR\right)\right|\left[J_{l-1}\left(\alpha_{lj'}x\right)J_{l}\left(\alpha_{lj}x\right)
\right],
\nonumber \\
I_{jj'}^{2} & =  \int_{0}^{1}xdx\left|\Delta\left(xR\right)\right|\left[J_{l}\left(\alpha_{lj'}x\right)J_{l+1}\left(\alpha_{lj}x\right)
\right], 
\end{align}
such that
\begin{equation}
\left|\Delta\right|_{l,jj'}^{++}=\frac{\left[I_{jj'}^{1}+I_{jj'}^{2}\right]}{\sqrt{\mathcal{N}_{j}\mathcal{N}_{j'}}}
\end{equation}
in the proper range of $j$ values.

\begin{figure}
\begin{centering}
\includegraphics[width=1\linewidth]{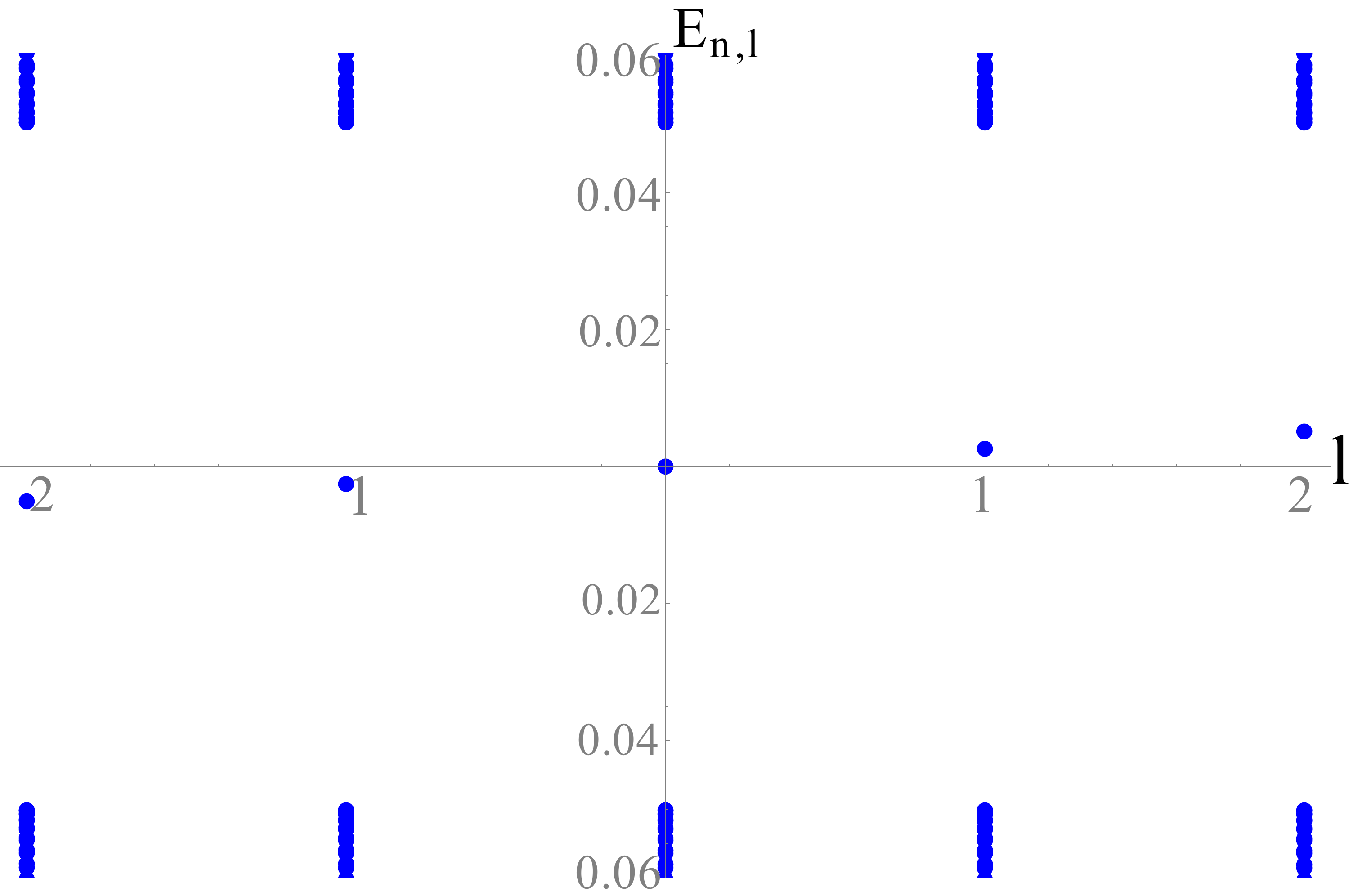}

\par\end{centering}

\protect\caption{Energy spectrum of a superconducting Weyl fermion with a vortex along the z-direction and with z-momentum $k_z=0$. Energy levels are labeled by a conserved angular momentum quantum number $l$ (states from $l=-2,...,2$ are shown); each value of angular momentum comprises a tower of scattering states with energy of the order of the gap $\Delta_0=0.05$ (parameters for our simulation follow in the main text) and an in-gap bound mode. The spectrum of the bound modes obeys roughly $E_{l}=\frac{\Delta_{0}^{2}}{E_{F}}l$, as expected.
is reasonably decent. A set of in-gap edge modes, localized at the surface of the cylinder also exists; their energies are omitted for the sake of clarity. \label{fig:numericspec}}
\end{figure}

We fix
\begin{align}
&
R=1000, \quad  N_{i}=450, \quad N_{0} =120, \quad \alpha=0.05,
\nonumber \\
&E_F=1, \quad m=1, \quad \Delta_{0}=0.05. 
\end{align}
The obtained spectrum follows in Fig.\ref{fig:numericspec}. The bound states have a spectrum with a linear dependence in the angular momentum $l$ and scale with the mini-gap $\frac{\Delta_{0}^{2}}{E_{F}}$. Quite generally, one may write $E_{l}=\frac{\Delta_{0}^{2}}{E_{F}}l$, for the vortex-bound states; of crucial importance is the absence of the half-integer shift in $l$, present in the Caroli-de Gennes-Matricon modes of vortices in s-wave superconductors in quadratically dispersing bands \cite{CdG}. It is precisely the $l=0$ zero-energy mode which will disperse linearly with $k_z$, forming the vortex-bound chiral modes.

\subsection*{z-dispersion and EM Hamiltonian}

Now we treat the Hamiltonians leading to dispersion along the vortex and the coupling to electromagnetic probes, $h_z$ and $h_{EM}$, respectively.
We make use of the states computed above to rewrite the these Hamiltonians in a basis which diagonalizes $h_{\perp}$, so that we can extract the $k_z$ dispersion, as well as electromagnetic coupling, of the bound states sector.

We will consider $\mathbf{A}=\left(0,0,Et\right)$, which is enough to probe for the electric charge of the vortex dispersing modes. Conveniently, as this field is independent of $z$, $k_{z}$ remains a good quantum number in this situation.

The fermionic field expansion in a complete basis now reads,
\begin{equation}
\Psi=\sum_{k_{z},n,l}e^{ik_{z}z}\phi_{nl}\left(r,\theta\right)f_{k_{z}nl}.
\end{equation}
We have
\begin{eqnarray}
\phi_{nl}\left(r\right) & = & \left(\begin{array}{c}
\int dk\left[c_{lk}^{n+}\chi_{kl}^{+}+c_{lk}^{n-}\chi_{kl}^{-}\right]\\
\int dk\left[d_{lk}^{n+}\tilde{\chi}_{kl}^{+}+d_{lk}^{n-}\tilde{\chi}_{kl}^{-}\right]
\end{array}\right)
\nonumber \\
 & \equiv & \left(\begin{array}{c}
\mathbf{u}_{ln}^{+}\left(r,\theta\right)+\mathbf{u}_{ln}^{-}\left(r,\theta\right)\\
\mathbf{v}_{ln}^{+}\left(r,\theta\right)+\mathbf{v}_{ln}^{-}\left(r,\theta\right)
\end{array}\right).
\end{eqnarray}
From our previous arguments, one might expect that $\mathbf{u}^{-}\approx\mathbf{v}^{-}\approx0$.
The complete basis for expansion, nevertheless, is this one and we use it for a complete expansion of the operator.

Breaking the full Hamiltonian as previously, $h_{\perp}+h_{z}+h_{EM}$, we have
\begin{eqnarray}
H & = & \sum_{k_{z},n,l}f_{k_{z}nl}^{\dagger}f_{k_{z}nl}E_{nl}
\nonumber \\
 &  & +\sum_{k_{z},n,l}\sum_{n',l'}f_{k_{z}n'l'}^{\dagger}\tilde{h}_{z}^{n'l',nl}f_{k_{z}nl}
 \nonumber \\
 &  & +\sum_{k_{z},n,l}\sum_{n',l'}f_{k_{z}n'l'}^{\dagger}\tilde{h}_{EM}^{n'l',nl}f_{k_{z}nl},
\end{eqnarray}
where $E_{nl}$ are the energies of the states at $k_z=0$ (out of which one separate sets $n_{scatt}$ and $n_{CdG}$ for the scattering and vortex bound states), and
\begin{align}
\tilde{h}_{z}^{n'l',nl} & =  v_{F}k_{z}\int d^{2}r\phi_{n'l'}^{\dagger}\left(r,\theta\right)\Gamma_{z}\phi_{nl}\left(r,\theta\right),
\nonumber \\
\tilde{h}_{EM}^{n'l',nl} & =  -v_{F}A_{z}\int d^{2}r\phi_{n'l'}^{\dagger}\left(r,\theta\right)\rho_{z}\Gamma_{z}\phi_{nl}\left(r,\theta\right).
\end{align}

Let us compute these matrix elements explicitly. First, notice that
\begin{align}
&\phi_{nl}\left(r,\theta\right)=\int dke^{i\left(l-\frac{\rho_{z}+\sigma_{z}}{2}\right)\theta}
\nonumber \\
 & \quad \times \left(\begin{array}{c}
\frac{c_{lk}^{n+}}{\sqrt{\mathcal{N}_{k}}}\left(\begin{array}{c}
J_{l-1}\left(kr\right)\\
iJ_{l}\left(kr\right)
\end{array}\right)+\frac{c_{lk}^{n-}}{\sqrt{\mathcal{N}_{k}}}\left(\begin{array}{c}
J_{l-1}\left(kr\right)\\
-iJ_{l}\left(kr\right)
\end{array}\right)\\
\frac{d_{lk}^{n+}}{\sqrt{\mathcal{N}_{k}}}\left(\begin{array}{c}
J_{l}\left(kr\right)\\
iJ_{l+1}\left(kr\right)
\end{array}\right)+\frac{d_{lk}^{n-}}{\sqrt{\mathcal{N}_{k}}}\left(\begin{array}{c}
J_{l}\left(kr\right)\\
-iJ_{l+1}\left(kr\right)
\end{array}\right)
\end{array}\right) \nonumber,
\end{align}
and that the exponential factor $e^{i\left(l-\frac{\rho_{z}+\sigma_{z}}{2}\right)\theta}$
commutes with both $\Gamma_{z}$ and $\rho_{z}\Gamma_{z}$. We then write
\begin{equation}
\phi_{nl}\left(r,\theta\right)=e^{i\left(l-\frac{\rho_{z}+\sigma_{z}}{2}\right)\theta}\tilde{\phi}_{nl}\left(r\right)
\end{equation}
and notice that
\begin{equation}
\left\langle nl|\Gamma_{z}|n'l'\right\rangle =\delta_{ll'}\int rdr\tilde{\phi}_{n'l}^{\dagger}\left(r\right)\Gamma_{z}\tilde{\phi}_{nl}\left(r\right)
\end{equation}
and similarly for $\rho_{z}\Gamma_{z}$. Making use of some continuum Bessel function identities, one obtains,
\begin{eqnarray}
\left\langle nl|\Gamma_{z}|n'l\right\rangle  & = & 2\int dk\left[c_{lk}^{n'+}c_{lk}^{n-}+c_{lk}^{n'-}c_{lk}^{n+}\right]\nonumber \\
 &  & -2\int dk\left[d_{lk}^{n'+}d_{lk}^{n-}+d_{lk}^{n'-}d_{lk}^{n+}\right] \nonumber \\
 & \equiv & \tilde{v}_{l}^{nn'}\\
\left\langle nl|\rho_{z}\Gamma_{z}|n'l\right\rangle  & = & 2\int dk\left[c_{lk}^{n'+}c_{lk}^{n-}+c_{lk}^{n'-}c_{lk}^{n+}\right]\nonumber \\
 &  & +2\int dk\left[d_{lk}^{n'+}d_{lk}^{n-}+d_{lk}^{n'-}d_{lk}^{n+}\right] \nonumber \\
 & \equiv & \tilde{q}_{l}^{nn'}.
\end{eqnarray}
These represent renormalizations of the velocity and electromagnetic charge felt by the z-dispersing modes in the Weyl superconductor with a vortex.
In this new basis we thus have
\begin{equation}
H=\sum_{k_{z},l,n,n'}f_{k_{z}ln}^{\dagger}\left[E_{nl}\delta_{nn'}+v_{F}\left(k_{z}\tilde{v}_{l}^{nn'}-\tilde{q}_{l}^{nn'}A_{z}\right)\right]f_{k_{z}ln'}.
\end{equation}

\subsection*{l=0 bound modes are chargeless}

Here it is our goal to show that $\tilde{q}_{0}^{nn'}=0$  if  $n$ and $n'$ correspond to bound states, which we write as $n_{CdG}$.

We start relating the coefficients $c^{\pm}$ and $d^{\pm}$. From particle-hole symmetry $\Xi=\rho_{y}\sigma_{y}K$.  Now $\left\{ L_{z},\Xi\right\} =0$ implies
\begin{eqnarray}
\Xi\psi_{nl}  & \propto & \psi_{n',-l}.
\end{eqnarray}

Using $J_{-l}=\left(-1\right)^{l}J_{l}$ for $l$ integer and that the coefficients $c^{\pm}$ and $d^{\pm}$ are real, the constant of proportion
is $-i$ and
\begin{eqnarray}
d_{lk}^{n'+} & = & c_{-lk}^{n+}\left(-1\right)^{l}\\
d_{lk}^{n'-} & = & -c_{-lk}^{n-}\left(-1\right)^{l}
\end{eqnarray}
with matching signs.

As the vortex bound modes decouple from the scattering ones, one expects that
\begin{eqnarray}
d_{lk}^{n_{CdG}+} & = & c_{-lk}^{n_{CdG}+}\left(-1\right)^{l}\\
d_{lk}^{n_{CdG}-} & = & -c_{-lk}^{n_{CdG}-}\left(-1\right)^{l},
\end{eqnarray}
which can be actually confirmed from our explicit numerically obtained coefficients, in the case of a finite cylinder.
Then, for the CdG modes, (for these we omit the $n$ labels)
\begin{widetext}
\begin{equation}
\psi_{n_{CdG}l}\left(r,\theta\right)=\int dk\frac{1}{\sqrt{\mathcal{N}_{k}}}\left(\begin{array}{c}
c_{lk}^{+}\left(\begin{array}{c}
e^{i\left(l-1\right)\theta}J_{l-1}\left(kr\right)\\
ie^{il\theta}J_{l}\left(kr\right)
\end{array}\right)+c_{lk}^{-}\left(\begin{array}{c}
e^{i\left(l-1\right)\theta}J_{l-1}\left(kr\right)\\
-ie^{il\theta}J_{l}\left(kr\right)
\end{array}\right)\\
c_{-lk}^{+}\left(-1\right)^{l}\left(\begin{array}{c}
e^{il\theta}J_{l}\left(kr\right)\\
ie^{i\left(l+1\right)\theta}J_{l+1}\left(kr\right)
\end{array}\right)-c_{-lk}^{-}\left(-1\right)^{l}\left(\begin{array}{c}
e^{il\theta}J_{l}\left(kr\right)\\
-ie^{i\left(l+1\right)\theta}J_{l+1}\left(kr\right)
\end{array}\right)
\end{array}\right)
\end{equation}
\end{widetext}
and, in particular, the $l=0$ mode reads
\begin{equation}
\psi_{n_{CdG}0}\left(r,\theta\right)=\int\frac{dk}{\sqrt{\mathcal{N}_{k}}}\left(\begin{array}{c}
\left(\begin{array}{c}
\left(c_{0k}^{+}+c_{0k}^{-}\right)e^{-i\theta}J_{1}\left(kr\right)\\
\left(c_{0k}^{+}-c_{0k}^{-}\right)iJ_{0}\left(kr\right)
\end{array}\right)\\
\left(\begin{array}{c}
\left(c_{0k}^{+}-c_{0k}^{-}\right)J_{0}\left(kr\right)\\
\left(c_{0k}^{+}+c_{0k}^{-}\right)ie^{i\theta}J_{1}\left(kr\right)
\end{array}\right)
\end{array}\right).
\end{equation}
In this case,
\begin{eqnarray}
 \int rdr\psi_{n_{CdG}0}^{\dagger}\Gamma_{z}\psi_{n_{CdG}0} & = & 4\int dkc_{0k}^{+}c_{0k}^{-}
\end{eqnarray}
and, similarly,
\begin{eqnarray}
 \int rdr\psi_{n_{CdG}0}^{\dagger}\rho_{z}\Gamma_{z}\psi_{n_{CdG}0} & = & 0.
\end{eqnarray}
Thus $\tilde{q}_{0}^{nn'}=0$  and the $l=0$ modes do not couple to electromagnetic fields, although they do disperse along $k_z$.

\subsection*{Vortex bound states spectrum is chiral}

Let us now compute the spectrum as function of $k_z$ perturbatively.
We neglect the effects of the vector potential and write
\begin{equation}
H=\sum_{k_{z},l,n,n'}f_{k_{z}ln}^{\dagger}\left[ h_{l}^{nn'}\left(k_{z}\right) \right]f_{k_{z}ln'}
\end{equation}
with the first quantized Hamiltonian
\begin{equation}
h_{l}\left(k_{z}\right)=E_{nl}\delta_{nn'}+v_{F}\tilde{v}^{nn'}_{l}k_{z}.
\end{equation}
Now we are able to compute the corrections of $k_{z}$ in the energy $E_{nl}$
at fixed $n=n_{CdG}$ and given $l$ in perturbation theory.

We start focusing in the subspace of CdG modes
\begin{equation}
h_{l}^{0}\left(k_{z}=0\right)=E_{n_{CdG}l}=l\times\frac{\Delta_{0}^{2}}{E_{F}}\equiv\delta_{l}.
\end{equation}
These states are non-degenerate in $l$ and perturbation theory is
computed easily. The first order corrections in $k_{z}$ read simply
\begin{equation}
E_{n_{CdG}l}^{(1)}=v_{F}\tilde{v}_{n_{CdG}n_{CdG}}^{l}k_{z},
\end{equation}
where
\begin{eqnarray}
\tilde{v}_{n_{CdG}n_{CdG}}^{l} & = & 4\sum_{j}\left[c_{lj}^{n_{CdG}+}c_{lj}^{n_{CdG}-}-d_{lj}^{n_{CdG}+}d_{lj}^{n_{CdG}-}\right]
\nonumber \\
 & = & 4\sum_{j}\left[c_{lj}^{n_{CdG}+}c_{lj}^{n_{CdG}-}+c_{-lj}^{n_{CdG}+}c_{-lj}^{n_{CdG}-}\right].
\end{eqnarray}
Now, to second order, we have
\begin{equation}
E_{n_{CdG}l}^{(2)}=\left(v_{F}k_{z}\right)^{2}\sum_{n\neq n_{CdG}}\frac{\tilde{v}_{n_{CdG}n}^{l}\tilde{v}_{nn_{CdG}}^{l}}{E_{n_{CdG}l}-E_{nl}}.
\end{equation}
For a fixed angular momentum $l$, the states labeled by $n\neq n_{CdG}$
separate in states of positive energy and states of negative energy,
which we label as $n_{\pm}$, respectively. These states have energies
of the order of the SC gap $\Delta_{0}$. We may then write
\begin{eqnarray}
\frac{1}{2v_F^2 \tilde{ m}_{l}} & \equiv & \sum_{n\neq n_{CdG}}\frac{\tilde{v}_{n_{CdG}n}^{l}\tilde{v}_{nn_{CdG}}^{l}}{E_{n_{CdG}l}-E_{nl}}\nonumber \\
 & = & \sum_{n_{+}}\frac{\tilde{v}_{n_{CdG}n_{+}}^{l}\tilde{v}_{n_{+}n_{CdG}}^{l}}{E_{n_{CdG}l}-E_{n_{+}l}}\nonumber \\
 &  & +\sum_{n_{-}}\frac{\tilde{v}_{n_{CdG}n_{-}}^{l}\tilde{v}_{n_{-}n_{CdG}}^{l}}{E_{n_{CdG}l}-E_{n_{-}l}}.
\end{eqnarray}
Let us focus shortly at $\tilde{m}_{0}$. In this case, since particle-hole
symmetry sends $l=0 \to 0$, then we have pairs of modes which obey $E_{n_{-}l}=-E_{n_{+}l}$
and $\tilde{v}_{n_{+}n_{CdG}}^{l}=\tilde{v}_{n_{-}n_{CdG}}^{l}$ (up
to a sign factor which cancels in the multiplication by the conjugate
transpose.) Then 
\begin{eqnarray}
\frac{1}{2v_F^2 \tilde{m}_{0}} & = & -\sum_{n_{+}}\frac{\tilde{v}_{n_{CdG}n_{+}}^{l}\tilde{v}_{n_{+}n_{CdG}}^{l}}{E_{n_{+}l}}\nonumber \\
 &  & -\sum_{n_{-}}\frac{\tilde{v}_{n_{CdG}n_{-}}^{l}\tilde{v}_{n_{-}n_{CdG}}^{l}}{E_{n_{-}l}}\nonumber \\
 & = & -\sum_{n_{+}}\frac{\tilde{v}_{n_{CdG}n_{+}}^{l}\tilde{v}_{n_{+}n_{CdG}}^{l}}{E_{n_{+}l}}\nonumber \\
 &  & +\sum_{n_{-}}\frac{\tilde{v}_{n_{CdG}n_{+}}^{l}\tilde{v}_{n_{+}n_{CdG}}^{l}}{E_{n_{+}l}}=0.
\end{eqnarray}
The $l=0$ states receive no contribution to second order in perturbation
theory. 

Let us make also a crude analysis of the energy denominator. We may
write the contributions approximately as
\begin{equation}
\frac{1}{\epsilon-\Delta_{0}}+\frac{1}{\epsilon+\Delta_{0}}\approx\frac{2\epsilon}{\Delta_{0}^{2}},
\end{equation}
where $\epsilon$ represents the mini-gap energies of the CdG modes
and$\Delta_{0}$ represents the energies of the gapped higher energy
modes. We clearly see that the sign of $\tilde{m}_{l}$ depends on the sign
of the CdG mode energies. Thus an analysis of the energy denominator
also suggests that if $l>0$ then $\tilde{m}_{l}>0$ and if $l<0$ then $\tilde{m}_{l}<0$.

So we may write
\begin{eqnarray}
E_{n_{CdG}l} & \approx & \delta_{l}+\tilde{v}_{n_{CdG}n_{CdG}}^{l}\left[v_{F}k_{z}\right]+\frac{\left[k_{z}\right]^{2}}{2\tilde{m}_{l}}.
\end{eqnarray}

Notice that for the charge elements $\tilde{q}_{l}$, the arguments
above follow much in the same way. The the effects of external fields in the $l=0$ mode also vanish to second order in perturbation theory, at least if one only considers couplings between the vortex bound states and the bulk scattering ones (that is, neglect the surface modes, as we have up to now).

We may thus extrapolate our perturbation power series as a Taylor series for a hyperbolic dispersion. Taking into account the signs of masses
and charges and the zero mode neutrality, while neglecting the system dependent quantities, a phenomenological Hamiltonian capturing the general physics of vortex bound modes in an infinite radius Weyl superconducting s-wave cylinder reads
\begin{eqnarray}
\tilde{h}_{z} & = & \int dk_{z}\gamma_{-k_{z}0} k_{z} \gamma_{k_{z}0}
\nonumber \\
 &  & +\sum_{l>0}\int dk_{z}c_{k_{z}l}^{\dagger}c_{k_{z}l}\sqrt{\left(k_{z}+Et\right)^{2}+\delta_{l}^{2}}
 \nonumber \\
 &  & -\sum_{l<0}\int dk_{z}c_{k_{z}l}^{\dagger}c_{k_{z}l}\sqrt{\left(k_{z}-Et\right)^{2}+\delta_{l}^{2}},
\end{eqnarray}
where $\gamma_{k_{z}0}$ are Majorana fermionic operators and $c_{k_{z}l}$ are regular complex fermions. The corresponding Lagrangian for the chiral modes, as used in the main text, follows from the standard procedure.

\section{QSHJJ pumps} \label{app:QSHJJapp}

The ground state of a Josephson junction at the edge of a quantum spin-Hall insulator evolves in a highly non-trivial way under an adiabatic ramp of the phase difference across the junction. In the presence of interactions, 
it was shown in Ref.\ [\onlinecite{ZK_PhysRevLett.113.036401}] that, despite the $2\pi$ periodicity of the Hamiltonian, an $8\pi$ evolution of the Josephson phase must occur before the junction's ground state returned to its original state. This anomalous $\mathbb{Z}_4$ Josephson effect has important consequences in the linking properties of chiral strings as it enforces that a process of ``tying'' a total of 8 links between the loops can actually be ``untied''  back into a pair of disjoint loops. Here we will revisit the physics of the  anomalous $\mathbb{Z}_4$ Josephson effect at the edges of quantum spin-Hall insulators, adding some details to previously known results, and expanding them to our present setting and purposes. 
We will follow Ref.\ [\onlinecite{ZK_PhysRevLett.113.036401}] and first treat the problem at the non-interacting level, then take into account density-density interactions inside the junction perturbatively; following that, we will introduce a bosonization picture to capture more details of the ground state evolution, like the pumping of fractionally charged quasi-particle.

\subsubsection{Microscopic fermion picture}

In our gauge choice, the present problem is equivalent both physically and algebraically to that of 
Ref.\ [\onlinecite{ZK_PhysRevLett.113.036401}] in the long junction limit. Let us start by diagonalizing the superconducting Hamiltonian \eqref{eq:JJBdG}. To achieve this, the bulk modes are treated simply as free right and left movers and we apply explicitly the superconducting boundary conditions at the edges
\begin{eqnarray}
\psi_{R}\left(0\right)&=&-ie^{-i\varphi}\psi_{L}^{\dagger}\left(0\right)\\\psi_{R}\left(L\right)&=&i\psi_{L}^{\dagger}\left(L\right).
\end{eqnarray}
As a result, the fermionic field operators can be expanded in two distinct sets $\sigma$ each with a tower of states labeled by an integer $n$,
\begin{eqnarray}
\Psi &=&\sum_{n,\sigma}\phi_{n,\sigma}c_{n\sigma}\\
c_{n\sigma}&=&\int dx\phi_{n,\sigma}^{\dagger}\Psi,
\end{eqnarray} 
where
\begin{eqnarray}
\phi_{n,+}&=&\frac{1}{\sqrt{2L}}\left(\begin{array}{c}
e^{iE_{n,+}x/v}\\
0\\
-ie^{i\frac{\varphi}{2}}e^{-iE_{n,+}x/v}\\
0
\end{array}\right)=\left(\begin{array}{c}
u_{n+}\\
0\\
v_{n+}\\
0
\end{array}\right)\\\phi_{n,-}&=&\frac{1}{\sqrt{2L}}\left(\begin{array}{c}
0\\
e^{-iE_{n,-}x/v}\\
0\\
ie^{i\frac{\varphi}{2}}e^{iE_{n,-}x/v}
\end{array}\right)=\left(\begin{array}{c}
0\\
u_{n-}\\
0\\
v_{n-}
\end{array}\right)
\end{eqnarray}
and where the single-particle energies are
\begin{equation}
E_{n\sigma}\left(\varphi\right)=\frac{v}{2L}\left(\pi (2n-\sigma)+\sigma\varphi\right).
\end{equation}
Notice also the $4\pi$ periodicity of the states under $\varphi$ evolution, characteristic of the $4\pi$ Josephson Effect in the presence of Majorana fermions.

It is useful to consider the discrete symmetries of the Hamiltonian. They are defined as
\begin{eqnarray}
PH:\,&&\rho_{y}\tau_{y}h\left(\varphi\right)^{*}\rho_{y}\tau_{y}=-h\left(\varphi\right),
\\TR:\,&&\tau_{y}h\left(\varphi\right)^{*}\tau_{y}=h\left(-\varphi\right).
\end{eqnarray}
Together with the periodicity of the Hamiltonian $h\left(\varphi\right)=h\left(\varphi+2\pi\right)$, the time-reversal operation enforces that $\varphi=0,\pi$ are two time-reversal invariant points, with a Kramers degenerate spectrum. The particle-hole constraint, which must be satisfied at every value of $\varphi$, fixes
\begin{eqnarray}
\rho_{y}\tau_{y}\phi_{n,\sigma}&=&-ie^{i\varphi/2}\phi_{-n,-\sigma}^{*},\\
c_{n,\sigma}^{\dagger}&=&-ie^{i\varphi/2}c_{-n,-\sigma},
\end{eqnarray}
at the first and second-quantized level, respectively. The application of particle-hole and time-reversal together enforces a chiral symmetry in the time-reversal invariant points which relates energies as $E_{-n}=-E_{n}$.

\begin{figure}[t!]
\begin{centering}
\includegraphics[width=1. \linewidth]{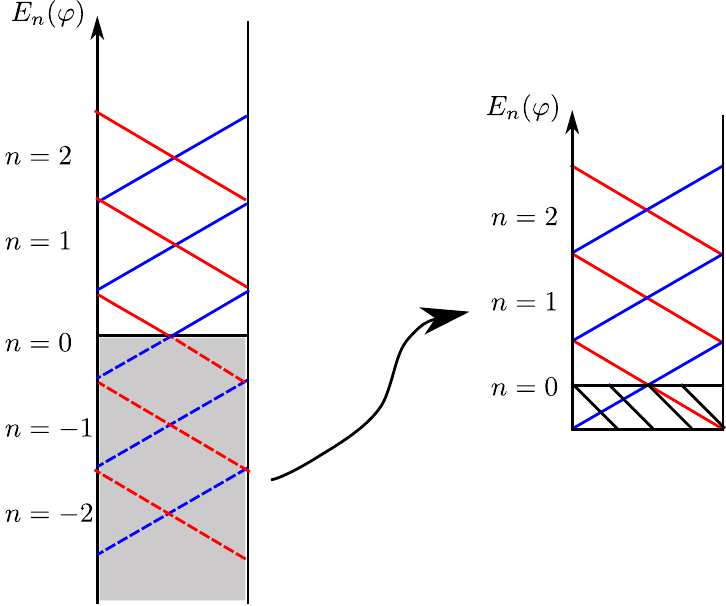}

\par\end{centering}
\protect\caption{Single particle spectrum analysis of the chiral loop linking problem. We measure energies in units of  $v/2L$ and plot them as function of $\varphi=0\to2\pi$, whose axis label is suppressed. The negative energy states are artificial from BdG doubling and are folded into positive energies by considering a filled sea of them, normal ordering the Hamiltonian. Blue and red correspond to $\sigma=\pm$, respectively. Depending if $\varphi<\pi$ or $\varphi>\pi$, the ground state will have a different filling.} \label{fig:spectrum_physical}
\end{figure}

The analysis of the particle-hole redundancy above implies that the physical Hilbert space is achieved after normal ordering with respect to the filled negative energy ground state as in Fig.\ \ref{fig:spectrum_physical}, obtaining
\begin{equation}
H=\sum_{\sigma}\sum_{n}\left(E_{n,\sigma}\left(\varphi\right)c_{n,\sigma}^{\dagger}c_{n,\sigma}\right).
\end{equation}

The normal ordering and building-up of the many body-spectrum here deserves some more in-depth attention, due to its non-trivial behavior upon variation of $\varphi$. From the single body energy spectrum, one sees that $E_{0,+}\left(\varphi\right)=-\pi+\varphi$ and $E_{0,-}\left(\varphi\right)=\pi-\varphi.$ For $\varphi<\pi$, the first state above has negative energy, while for $\varphi>\pi$ it is the second one that has, in fact, negative energy. We normal order with respect to a negative-energy filled ground state. If we classify the states by the parity of occupation of the state created by $c_{0,+}^{\dagger}$, that is $\mu=\left(-1\right)^{c_{0,+}^{\dagger}c_{0,+}}$, then
\begin{eqnarray}
&&\left|GS\right\rangle =\left|\mu=1\right\rangle= c_{0,+}^{\dagger}\prod_{\sigma,n<0}c_{n,\sigma}^{\dagger}\left|0\right\rangle,\,\varphi<\pi\\ 
&&\left|GS\right\rangle =\left|\mu=0\right\rangle=  c_{0,-}^{\dagger}\prod_{\sigma,n<0}c_{n,\sigma}^{\dagger}\left|0\right\rangle ,\,\varphi>\pi.
\end{eqnarray}
where $\left|0\right\rangle $ is the empty state. The ground state parity is phase dependent. This is the hallmark of the $4\pi$ periodic Josephson-effect.

We can proceed with this analysis and build the many-body spectrum, classifying it according to the parity of each state. Above the ground states, the first excited states, for example, are readily written
\begin{eqnarray}
\left|1\right\rangle 	&=&	c_{0,-}^{\dagger}\left|\mu=1\right\rangle ,\,\varphi<\pi\\
\left|1\right\rangle 	&=&	c_{0,+}^{\dagger}\left|\mu=0\right\rangle ,\,\varphi>\pi.
\end{eqnarray}
Through the relationship $c_{0,-}^{\dagger}\sim c_{0,+}$ we see that the first excited states have opposite parity with respect to the ground state.

\begin{figure}[t!]
\begin{centering}
\includegraphics[width=1. \linewidth]{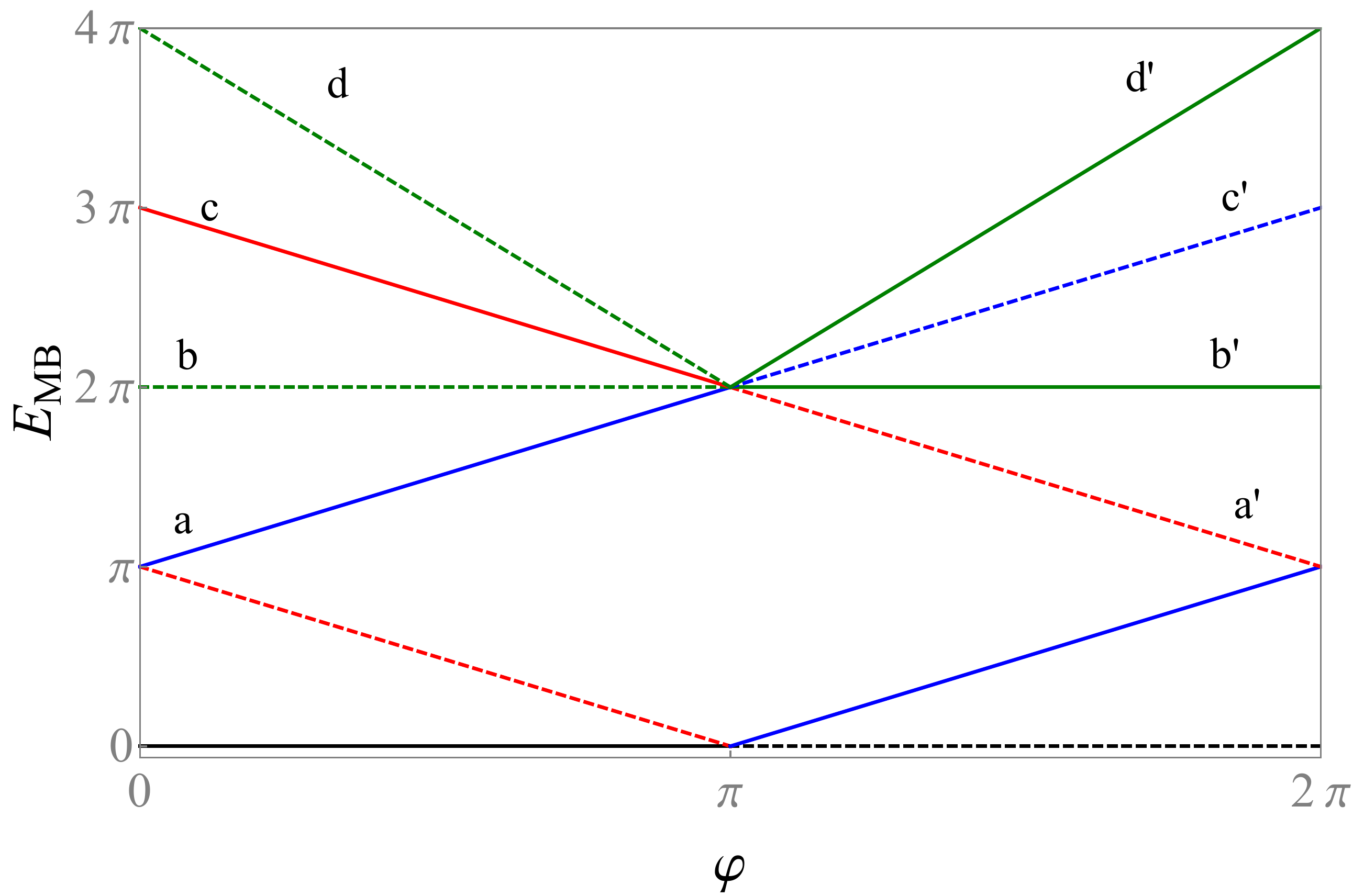}

\par\end{centering}
\protect\caption{Many-body spectrum of the chiral loop linking problem. Some relevant states of single (blue and red) and double (dashed green) occupancy are presented. Notice the fourfold degeneracy at $\varphi=\pi$. The four degenerate states are identified by their creation operators applied to the vacuum (black horizontal line). Interactions can open up this fourfold degeneracy without breaking the time-reversal invariance of the point.} \label{fig:MBspectrum}
\end{figure}

It will prove relevant to study also the next four excited levels. They will comprise a set of four states which become degenerate at $\varphi=\pi$. Starting with $\varphi<\pi$, they are given by
\begin{eqnarray}
\left|2a\right\rangle 	&=&	c_{1,+}^{\dagger}\left|\mu=1\right\rangle =c_{1,+}^{\dagger}\left|GS\right\rangle  \\
\left|2b\right\rangle 	&=&	c_{1,+}^{\dagger}c_{0,-}^{\dagger}\left|\mu=1\right\rangle \sim c_{1,+}^{\dagger}c_{0,+}\left|GS\right\rangle \\
\left|2c\right\rangle 	&=&	c_{1,-}^{\dagger}\left|\mu=1\right\rangle =c_{1,-}^{\dagger}\left|GS\right\rangle \\
\left|2d\right\rangle 	&=&	c_{1,-}^{\dagger}c_{0,-}^{\dagger}\left|\mu=1\right\rangle \sim c_{1,-}^{\dagger}c_{0,+}\left|GS\right\rangle .
\end{eqnarray}
For $\varphi>\pi$, the ground state changes. States like 2b are not possible anymore, as one cannot remove particles with $c_{0+}$, or the full vector vanishes. The new set of states is, from lowest to largest energy, in fact given by
\begin{eqnarray}
\left|2a'\right\rangle &=&	c_{1,-}^{\dagger}\left|\mu=0\right\rangle =c_{1,-}^{\dagger}\left|GS\right\rangle \\
\left|2b'\right\rangle 	&=&	c_{1,-}^{\dagger}c_{0,+}^{\dagger}\left|\mu=0\right\rangle \sim c_{1,-}^{\dagger}c_{0,-}\left|GS\right\rangle \\
\left|2c'\right\rangle 	&=&	c_{1,+}^{\dagger}\left|\mu=0\right\rangle =c_{1,+}^{\dagger}\left|GS\right\rangle \\
\left|2d'\right\rangle 	&=&	c_{1,+}^{\dagger}c_{0,+}^{\dagger}\left|\mu=0\right\rangle \sim c_{1,-}^{\dagger}c_{0,-}\left|GS\right\rangle .
\end{eqnarray} 
The proper application of the $c_{0,\pm}$ operators in the ground state allows one to compute the parity of each of the excited ones. The corresponding many-body spectrum is produced in Fig.\ \ref{fig:MBspectrum}, with the parity of each state represented by full/dashed lines, showing that the four lowest-energy two-particle states are degenerate at $\varphi=\pi$. These results match, with those of Ref.\ [\onlinecite{ZK_PhysRevLett.113.036401}] in the long-junction limit.

As previously discussed, the periodicity of the ground state evolution in $\varphi$ is $4\pi$, in contrast with the Hamiltonian which is $2\pi$ periodic, and we have the regular $\mathbb{Z}_2$ parity pump of topological Josephson junctions. Yet, this is a single-body scenario. It is a remarkable property of quantum spin-Hall Josephson junctions that interactions may lead to more complex fractionalization of the Josephson effect, even without breaking the time-reversal symmetry which is responsible, in part, for the state degeneracies. In fact, while time-reversal symmetry protects the degeneracies at $\varphi=0,\, \pi$ by Kramers' theorem, one also has the further degeneracy due to the ground state parity $\mu=\left(-1\right)^{c_{0,+}^{\dagger}c_{0,+}}$. This can then be lifted, for example, by density-density interactions, which penalize in energy states with more particles. 

Indeed, introducing an interaction Hamiltonian
\begin{equation}
H_{I}=\lambda\int_{0}^{L}dx\,n^{2}
\end{equation}
with $n=\psi_{R}^{\dagger}\psi_{R}+\psi_{L}^{\dagger}\psi_{L}$, and treating its effects perturbatively on the four states at $\varphi=\pi$ (the calculations are involved and will omitted here), one obtains 
\begin{equation}
\left\langle \mu',\sigma'\left|:H_{I}:\right|\mu,\sigma\right\rangle=m_{z}\mu_{z}\sigma_{z},
\end{equation}
up to a global energy shift. The four degenerate states gap-out by
\begin{align}
&\quad 
m_{z}=\lambda\int_{0}^{L}dx\left|u_{0,+}^{*}u_{-1,-}-u_{1,+}^{*}u_{0,-}\right|^{2}
\nonumber \\
&\Rightarrow 2m_{z}=\frac{\lambda}{L},
\end{align}
which can be, again, compared to the result of Ref.\ [\onlinecite{ZK_PhysRevLett.113.036401}] in the large $L$ limit.

\begin{figure}[t!]
\begin{centering}
\includegraphics[width=1. \linewidth]{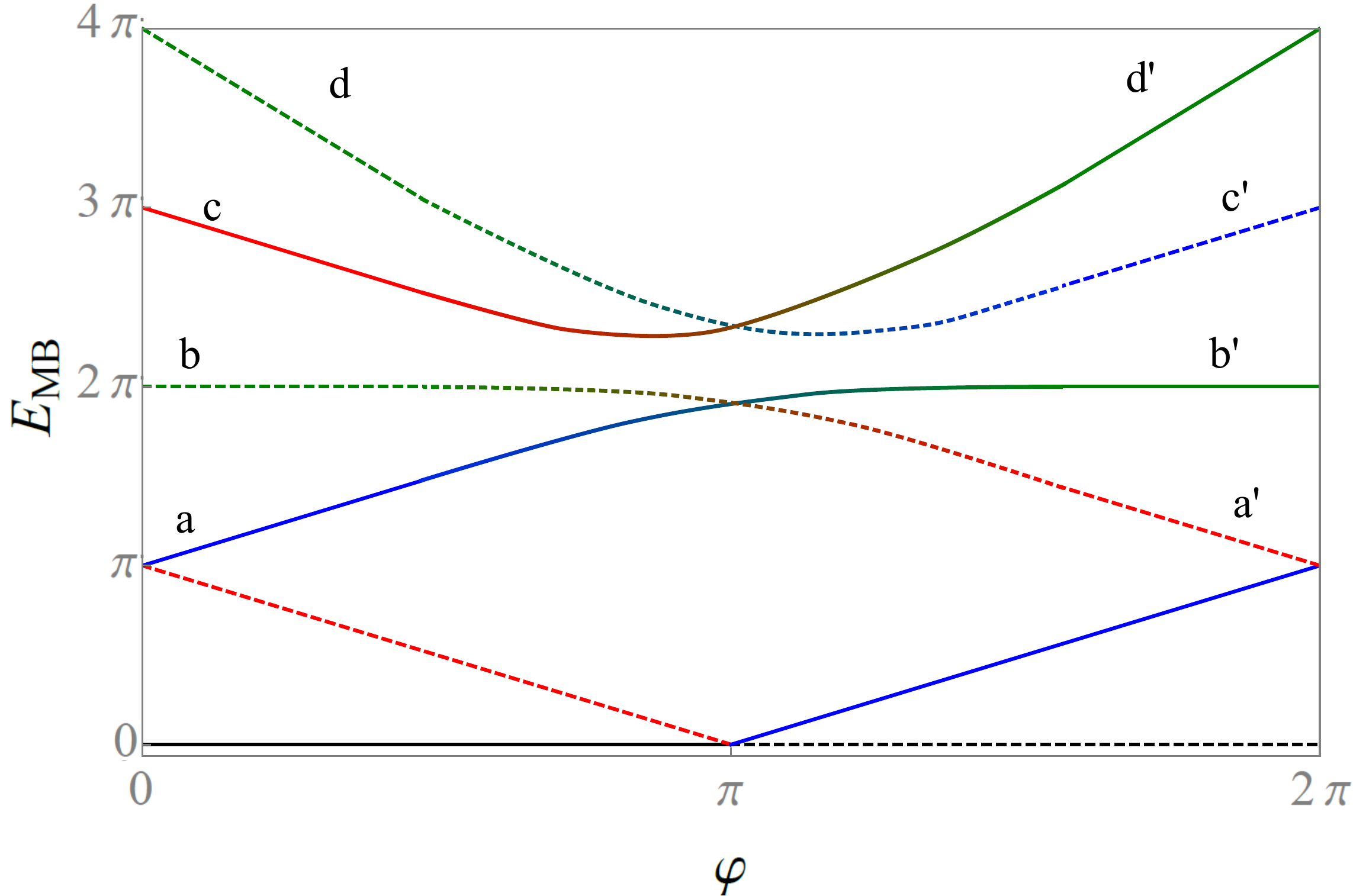}

\par\end{centering}
\protect\caption{Many-body spectrum of the chiral loop linking problem in the presence of density-density interactions. While crossings protected by time-reversal symmetry remain intact, the four-fold degeneracy at phase $\varphi=\pi$ is lifted by the interactions. The parity of the states is preserved upon gapping.} \label{fig:GapMBspec}
\end{figure}

In this many-body scenario, Kramers' symmetry is retained, as time-reversal symmetry is not broken. The realization of time-reversal symmetry in this problem connects sectors $\mu,\sigma$ to $-\mu,-\sigma$, such that the product $\mu\sigma$ is a good quantum number. The gap opens between $\mu\sigma=+1$ and $\mu\sigma=-1$ as represented in Fig.\ \ref{fig:GapMBspec}. Perturbations which break the time-reversal symmetry are also allowed, and would shift the avoided crossing region away from $\varphi=\pi$, yet, not changing the physical scenario of the $\mathbb{Z}_4$ Josephson effect.

\subsubsection{Bosonization picture}
In order to develop an understanding of the physical implications of the $8\pi$ periodic Josephson effect, or rather the periodicity of the string Hilbert-space rearranging under an 8-fold linking of two $(1,1)$ vortices, we can apply a bosonization analysis of the previous problem. We again follow and expand the analysis of Ref.\ [\onlinecite{ZK_PhysRevLett.113.036401}] in somewhat more detail, so that the results of this section serve both as a check of their result, as well as a pointing to, and explanation of, subtleties which deserve further discussion.
 
 Defining $\eta=1,-1$ for $R,L$ modes, respectively, one identifies
\begin{equation}
\psi_{\eta}=\frac{1}{\sqrt{2\pi x_{c}}}e^{i\phi_{\eta}},
\end{equation}
where $x_{c}$  regularizes the theory at large momentum and the bosonic fields obey equal time commutation relations
\begin{eqnarray}
\left[\phi_{\eta}\left(x\right),\phi_{\eta'}\left(x'\right)\right]
&=&
i\pi\delta_{\eta\eta'}\left(-1\right)^{\eta}sgn\left(x-x'\right)
\nonumber \\&&+i\pi sgn\left(\eta-\eta'\right).
\end{eqnarray}
 If one further introduces the conjugate fields
 \begin{eqnarray}
2\phi&=&\phi_{R}+\phi_{L}\\
2\theta&=&\phi_{R}-\phi_{L},
\end{eqnarray}
one can rewrite the non-interacting quantum-spin-Hall Josephson-junction Hamiltonian as
\begin{eqnarray}
H_{NI}&=&\int dx\frac{2v}{2\pi}\left[\left(\partial_{x}\theta\right)^{2}+\left(\partial_{x}\phi\right)^{2}\right]
\nonumber \\
&&+
U_0 \int dx
\left[\delta\left(x\right)\cos\left(2\phi-\frac{\pi}{2}+\varphi\right)\right. \nonumber\\
&& \quad 
\left.-\delta\left(x-L\right)\cos\left(2\phi-\frac{\pi}{2}\right)\right]
\end{eqnarray}
where $U_0=v/2\pi x_{c}$. In terms of the bosonic variables, the boundary conditions thus read
\begin{eqnarray}
\phi_{R}\left(0\right)+\phi_{L}\left(0\right)+\varphi&=&\frac{\pi}{2}+2n\pi \\
\phi_{R}\left(L\right)+\phi_{L}\left(L\right)&=&\frac{\pi}{2}+2n^{\prime}\pi, 
\end{eqnarray}
where $n,n'$ are fixed by the ground state. 

Now we move to interactions. The first interaction of our interest is the density-density term of before, which translates into
\begin{eqnarray}
H_{I}=\frac{\lambda}{\pi^{2}}\int dx\left(\partial_{x}\theta\right)^{2}.
\end{eqnarray}
This is, however, not the most interesting term from the bosonization point of view. Under time-reversal operations, the bosonic fields transform as $\phi\rightarrow -\phi$  and $\theta\rightarrow \theta+\pi$. Therefore, more sine-Gordon terms are possible. In particular, the pair-backscattering term
\begin{equation}
H_{PB}=\lambda'\int dx:\psi_{R}^{\dagger}\psi_{R}^{\dagger}\psi_{L}\psi_{L}+H.c.:
\end{equation} 
is time reversal symmetric and translates into a $\cos{4\theta}$ term under the bosonization dictionary.

 Putting all together, the full bosonic Hamiltonian density of interest in our problem reads
\begin{eqnarray}
\mathcal{H}=\mathcal{H}_0+\mathcal{H}_{\theta}+\mathcal{H}_{\phi}
\end{eqnarray}
where
\begin{eqnarray}
\mathcal{H}_0&=& 
\frac{\tilde{v}}{2\pi}\left[g\left(\partial_{x}\phi\right)^{2}+\frac{1}{g}\left(\partial_{x}\theta\right)^{2}\right], \label{eq:Hphitheta}\\
\mathcal{H}_{\phi}&=&
U_{0}\left[\delta\left(x\right)\cos\left(2\phi-\frac{\pi}{2}+\varphi\right)\right. \nonumber\\
&&\quad 
\left.-\delta\left(x-L\right)\cos\left(2\phi-\frac{\pi}{2}\right)\right],
\\
\mathcal{H}_{\theta}&=& V_{0}\cos4\theta.
\end{eqnarray}
The Fermi velocity here becomes $\tilde{v}=\left(2v\right)/g$ and is rescaled by the Luttinger parameter $g= \left(1+\frac{2\lambda}{\left(2v\right)\pi} \right)^{1/2}.$ (we will be carrying the unconventional factor of $2$ together with the velocity $v$. This is an artifact of our mapping between Majorana loops and the Josephson junction problem and will be used to keep track of the true parameters of the original loop problem.)

To extract the physics of this Hamiltonian, one first notices that $\phi$ and $\theta$ are conjugate. This means that $H_{\theta}$ and $H_{\phi}$ compete with one another.  We focus in the bulk of the problem, where $H_{\phi}=0$, and use the superconducting gap from the latter only to fix $\theta$ as a compact variable modulo $2\pi$, which reflects the condensation of Cooper pairs at the edges. Then, with only the bulk Hamiltonian, it is clear that we choose $\theta$ as a preferred variable over $\phi$; from this, one writes the Action
\begin{eqnarray}
S=\int d^{2}x\left[\frac{1}{2\pi g}\left(\partial\theta\right)^{2}+V_{0}\cos4\theta\right]
\end{eqnarray}
and extract the correlation function of the general vertex operators
\begin{eqnarray}
\left\langle e^{in\theta\left(0\right)}e^{-in\theta\left(x\right)}\right\rangle = \left(\frac{1}{\left|\mathbf{x}\right|}\right)^{n^{2}g/2}
\end{eqnarray}
and read the corresponding scaling dimension $\tilde{\Delta}_{n}=n^{2}\frac{g}{4}$.  Relevant operators obey $\tilde{\Delta}_{n}<2$, such that for our particular backscattering potential $n=4$ and
\begin{equation}
\frac{\lambda}{\left(2v\right)}>\frac{3\pi}{2}\sim4.71.
\end{equation}
This is the condition for which $\cos4\theta$ flows to strong coupling, pinning $\theta$ to discrete values throughout the junction. 

We may now restart the problem close to the fixed point of pinned $\theta$, treating the effects from $\mathcal{H}_{\phi}$ perturbatively.  The low-energy Hamiltonian will describe the time fluctuations of a single degree of freedom $\theta$, constant  through the junction length $x:0\rightarrow L$ and whose Hilbert space contains four distinct coherent states of fixed values
\begin{eqnarray}
\theta_{n}=\frac{\pi}{4}+\frac{\pi}{2}n,\ n=0,...,3. \label{eq:thetas}
\end{eqnarray}
The corresponding Hamiltonian can be read off from the densities \eqref{eq:Hphitheta}  with the $\theta$-conjugate momentum $\Pi=\frac{1}{\pi}\partial_{x}\phi$, from Hamilton's equations, and we find
\begin{equation}
H=\frac{\tilde{v}L}{2\pi}\left[g\pi^{2} \Pi^{2}\right]+V_{0}L\cos4\theta.
\end{equation}
Here  $\theta$  is an operator which measures the value of $\theta_{n}$  on the distinct states of fixed value. 

The boundary Hamiltonian is written
\begin{eqnarray}
V&=&\frac{U_{0}}{2}\left[e^{i\left(2\phi\left(0\right)-\frac{\pi}{2}+\varphi\right)}+e^{-i\left(2\phi\left(0\right)-\frac{\pi}{2}+\varphi\right)}\right.
\nonumber \\
&&\left.-e^{i\left(2\phi\left(L\right)-\frac{\pi}{2}\right)}-e^{-i\left(2\phi\left(L\right)-\frac{\pi}{2}\right)}\right]
\end{eqnarray}
and will then be treated in perturbation theory. For this, we need to fix the Hilbert space of the unperturbed problem. One must be careful, as a gauge freedom exists between the four states which must be accounted for, and is fixed by the superconducting phase $\varphi$ which appears in the perturbation Hamiltonian $V$. 

To fix this gauge freedom issue, we act in order to transfer all the superconducting phase dependence from the perturbation to the states. We start with the operators
\begin{eqnarray}
\hat{O}_{\alpha}&=&e^{i\alpha\theta},
\nonumber \\ 
\hat{Q}_{\beta}&=&e^{i\beta\int_{x_{1}}^{x_{2}}dx\Pi\left(t,x\right)}
=e^{i\frac{\beta}{\pi}\left(\phi\left(x_{1}\right)-\phi\left(x_{2}\right)\right)},
\end{eqnarray}
 whose actions on general bosonic coherent states are given by
 \begin{eqnarray}
 \hat{O}_{\alpha}\left|\theta\left(t,x\right)\right\rangle &=&e^{i\alpha\theta\left(t,x\right)}\left|\theta\left(t,x\right)\right\rangle, 
 \nonumber \\
 \hat{Q}_{\beta}\left|\theta\left(t,x\right)\right\rangle &=&\left|\theta\left(t,x\right)+\beta\theta\left(x-x_{1}\right)\theta\left(x_{2}-x\right)\right\rangle .
\end{eqnarray}
As $\Pi$ is the momentum conjugate to $\theta$, $\hat{Q}_{\beta}$ is a translation operator for $\theta$ by a quantity $\beta.$ The four distinct ground states contain constant fields through the chain distinguished by shifts of $\pi/2$ as seen in \eqref{eq:thetas}. Thus, this brings our attention to the operator
\begin{equation}
\hat{Q}_{\pi/2}=e^{i\frac{1}{2}\left(\phi\left(L\right)-\phi\left(0\right)\right)},
\end{equation}
which translates $\theta$ by $\pi/2$ globally in the junction. As a final step we  translate the bosonic field at an extremum as $\phi\left(0\right)\rightarrow\phi\left(0\right)-\frac{\varphi}{2}$. This may introduce a kink in the field through the junction, in order not to change the boundary condition at the other extremity, but this process can at worst introduce a constant energy shift to the Hamiltonian. As a result, the perturbation Hamiltonian simplifies to
\begin{eqnarray}
V&=&\frac{U_{0}}{2}\left[e^{i\left(2\phi\left(0\right)-\frac{\pi}{2}\right)}+e^{-i\left(2\phi\left(0\right)-\frac{\pi}{2}\right)}\right.
\nonumber \\
&&\left.-e^{i\left(2\phi\left(L\right)-\frac{\pi}{2}\right)}-e^{-i\left(2\phi\left(L\right)-\frac{\pi}{2}\right)}\right] \label{eq:perturb}
\end{eqnarray}
while the translation operator becomes
\begin{eqnarray}
\hat{Q}_{\pi/2}=e^{i\frac{\varphi}{4}}e^{i\frac{1}{2}\left(\phi\left(L\right)-\phi\left(0\right)\right)}.
\end{eqnarray}
The problem is ready. We have the perturbation Hamiltonian \eqref{eq:perturb} and a carefully built set of states in the Hilbert space
\begin{equation}
\left|n\right\rangle \equiv e^{in\frac{\varphi}{4}}\left|\theta_{n}\right\rangle =\hat{Q}_{\pi/2}^n\left|\theta_{0}\right\rangle,
\end{equation}
where $\left|\theta_{0}\right\rangle$ is a defined reference state. Under this procedure, the superconducting boundary term indeed just fixes the periodicity in $\theta$, which becomes compact, defined modulo $2\pi$, while the superconducting phase difference across the junction is carried only by the distinct states. 

We are ready to extract physical quantities out of this problem. First, we study the charge difference between adjacent states. We find
\begin{eqnarray}
&&\left\langle n+1\left|\int:n:\right|n+1\right\rangle -\left\langle n\left|\int:n:\right|n\right\rangle
\nonumber \\&=&\frac{1}{\pi}\left[\left\langle n+1\left|\theta\right|n+1\right\rangle -\left\langle n\left|\theta\right|n\right\rangle \right]
\nonumber \\
&=&\frac{1}{2},
\end{eqnarray}
or
\begin{equation}
Q_{n+1}-Q_{n}=\frac{e}{2}.
\end{equation}
Up to a global ground state charge, the four states  $\left|\theta_{n}\right\rangle$ have charge differences as multiples of $e/2$. 

One may then compute the transition amplitude of $\theta_{n}(t)$ between two states $ n $ and $n'$ as
\begin{align}
\left\langle n'\left|e^{-HT}\right|n\right\rangle 
&=
e^{i\left(n-n'\right)\frac{\varphi}{4}}\left\langle \theta_{n'}\left|e^{-HT}\right|\theta_{n}\right\rangle 
\nonumber \\
&\equiv
e^{i\left(n-n'\right)\frac{\varphi}{4}}A_{n'n},
\end{align}
where
\begin{equation}
A_{n'n} =\frac{1}{\mathcal{N}}\int\left[d\theta\right]e^{-S_{E}\left[\theta\right]}
\end{equation}
with Euclidean action
\begin{equation}
S_{E}\left[\theta\right]=\int_{0}^{T}dt\left[\frac{2\pi vL}{2}\left(\partial_{t}\theta\right)^{2}+V_{0}L\cos4\theta\right]
\end{equation}
and boundary conditions $\theta\left(0\right)=\theta_{n}$  and $\theta\left(T\right)=\theta_{n'}$.

The computation of the amplitude may be done in the standard instanton gas approximation \cite{CallanColeman} and shows an energy splitting between the four, otherwise degenerate, states, given by
\begin{equation}
\Delta\epsilon_{n',n}=Ke^{-S_{inst}^{n',n}}\equiv t_{n',n}.
\end{equation}
where the classical instanton action is $S_{inst}^{n',n}=\int_{\theta_{n}}^{\theta_{n'}}d\theta L\sqrt{4\pi vV_{0}\cos4\theta}$ and $K$  is a numerical constant that can be computed out of the functional determinant and is of the order of unity.

This energy splitting can be captured in an effective tight-binding Hamiltonian
\begin{equation}
H_{eff}=-\sum_{n,n'}t_{n,n'}e^{i\left(n-n'\right)\frac{\varphi}{4}}\left|\theta_{n}\right\rangle \left\langle \theta_{n'}\right|.
\end{equation}
From the instanton classical action, one sees that $\log\frac{t_{n,0}}{t_{n',0}}=L\sqrt{4\pi vV_{0}}(n'-n)I_{1}$, where $I_{1}=\int_{\pi/4}^{\pi/4+\pi/2}d\theta\sqrt{\cos4\theta}$ is an elliptic integral. This implies that the ratio of the hopping amplitudes between neighbors falls off exponentially with the distance between the neighbors. If only nearest and next-nearest neighbors are kept, the effective Hamiltonian reduces to that of Ref.\ [\onlinecite{ZK_PhysRevLett.113.036401}], which describes a dynamical hopping of $\theta$ between states transporting charge $e/2$ and $e$ in nearest and next-nearest neighbor hopping processes, respectively.

\section{The chiral $SO(N)_1$ current algebra}  \label{app:SONWZW}
Central to our discussions is the low-energy effective theory describing $N$ right-moving Majorana fermions along a vortex 
\begin{align}
\mathcal{L}=\sum_{a=1}^Ni\gamma^a_R(\partial_t-v\partial_x)\gamma^a_R.
\end{align}
It has an emergent global $SO(N)$ symmetry that rotates the Majorana fermions $\gamma^a\to\mathcal{O}^a_b\gamma^b$, where $\mathcal{O}^a_b$ is a $N\times N$ orthogonal matrix. Each vortex line is thus characterized by a $so(N)$ Wess-Zumino-Witten (WZW) theory\cite{WessZumino71,WittenWZW} or affine Kac-Moody algebra at level 1. Here we review some relevant features of the $so(N)_1$ algebra, which are well-known and can be found in standard texts on conformal field theory (CFT) such as Ref.\onlinecite{bigyellowbook}. 

The $so(N)_1$ currents have the free field representation
\begin{align}
J^\beta(z)=\frac{i}{2}\boldsymbol\gamma(z)^Tt^\beta\boldsymbol\gamma(z)=\frac{i}{2}\sum_{ab}\gamma^a(z)t^\beta_{ab}\gamma^b(z)\label{so(N)current}
\end{align} 
where the $t^\beta=\left(t^{\beta}_{ab}\right)_{N\times N}$'s are antisymmetric real matrices that generate the $so(N)$ Lie algebra and $z=e^{\tau+ix}$ is the complex space-time parameter along the vortex direction $x$; it is implicitly understood that \eqref{so(N)current} is normal ordered. The operator product expansion (OPE) for the Majorana fields
\begin{align}\gamma^a(z)\gamma^b(w)=\frac{\delta^{ab}}{z-w}+\ldots\label{fermionOPE}
\end{align} 
imply the $so(N)_1$ currents obey the OPE
\begin{align}J^\beta(z)J^\gamma(w)=\frac{\delta^{\beta\gamma}}{(z-w)^2}+\sum_\delta\frac{if_{\beta\gamma\delta}}{z-w}J^\delta(w)+\ldots\label{so(N)1Jrelation}
\end{align} 
where $f_{\beta\gamma\delta}$ are the structure constants of the $so(N)$ Lie algebra.

Since the $SO(N)_1$ WZW CFT describes a free chiral $N$-component Majorana fermion theory, it carries an anomalous heat (energy) current~\cite{KaneFisher97, Cappelli01, Luttinger64} that only propagate along a single direction at temperature $T$ 
\begin{align}
I_T\approx c_-\frac{\pi^2k_B^2}{6h}T^2,\label{thermalcurrent}
\end{align}
where $c_-=N/2$ is called the chiral central charge of the CFT. In general $c_-=(N_R-N_L)/2$ for a system of $N_R$ right-moving Majorana fermions and $N_L$ left-moving ones. Here we extend the $SO(N)_1$ WZW CFT to negative $N$ so that it describes $N$ right-moving Majorana fermions when $N$ is positive or $-N$ left-moving ones when $N$ is negative.

Excitations along the vortex line transform according to the $SO(N)$ symmetry. They decompose into {\em primary fields} and their corresponding descendants. A primary field ${\bf V}_\lambda=(V^1,\ldots,V^d)$ is a simple excitation sector that irreducibly represents the $so(N)_1$ Kac-Moody algebra, i.e.,
\begin{align}
J^\beta(z)V^r(w)=-\sum_{s=1}^d\frac{(t^\beta_\lambda)_{rs}}{z-w}V^s(w)+\ldots\label{currentrepOPE}
\end{align} 
where $\lambda$ labels some $d$-dimensional irreducible representation of $so(N)$ and $t^\beta_\lambda$ is the $d\times d$ matrix representing the generator $t^\beta$ of $so(N)$. For example, it is straightforward to check by using the definition \eqref{so(N)current} and the OPE \eqref{fermionOPE} that the Majorana fermion $\boldsymbol\gamma=(\gamma^1,\ldots,\gamma^N)$ is primary with respect to the fundamental vector representation of $SO(N)$, i.e. 
\begin{align}
J^\beta(z)\gamma^a(w)=-\sum_{b=1}^N\frac{t^\beta_{ab}}{z-w}\gamma^b(w)+\ldots.
\end{align} 
There are extra primary fields other than the the trivial vacuum $1$ and the fermion $\gamma$. The spinor representations $\sigma$, for $N$ odd, or $s_+$ and $s_-$, for $N$ even, also correspond to primary fields of $so(N)_1$. Their scaling dimensions ,or conformal spins, are 
\begin{align}
h_\sigma=\frac{N}{16},\quad h_{s\pm}=\frac{N}{16}.\label{spinsigmas}
\end{align} 
Unlike the infinite number of irreducible representations of a Lie algebra, the extended affine $so(N)_1$ algebra only has a truncated set of primary fields $\{1,\gamma,\sigma\}$, for $N$ odd, or $\{1,\gamma,s_+,s_-\}$, for $N$ even. 

To fully characterize a set of primary fields, besides their quantum numbers, one has to determine the so-called fusion rules. These come as a set of multiplication rules which determine which primary fields can appear in the decomposition of a product of other two. In our present case, the fusion rules read $\gamma\times\gamma=1$, 
\begin{align}
\sigma\times\gamma=\sigma,\quad\sigma\times\sigma=1+\gamma,
\end{align} 
for $N$ odd, 
\begin{align}
s_\pm\times\gamma=s_\mp
\end{align} for $N$ even, and 
\begin{align}
s_\pm\times s_\pm=1\quad\mbox{or}\quad\gamma
\end{align}
for $N\equiv0$ or 2 modulo 4 respectively. To obtain the fusion rules of a CFT, the easiest path it to consider its modular information, in this case the modular information of the $SO(N)_1$ WZW CFT. By modular information we refer to the $S$ and $T$ matrices that represent the $SL(2,\mathbb{Z})$ transformations when the CFT is put on a periodic torus spacetime geometry; they correspond to inversions followed reflections, and unit translations of the torus cycles, respectively. In our case, they are given $3\times 3$ or $4\times 4$ matrices, depending whether $N$ is odd or even, with entries labeled by the primary fields. The $T$ matrix can be chosen to be diagonal with entries given by the $2\pi$ twist phases of primary fields 
 \begin{align}
 T_{{\bf a}{\bf b}}=\delta_{{\bf a}{\bf b}}e^{2\pi ih_{\bf a}}.\label{Tmatrix}
 \end{align} 
 The $S$ matrix, corresponding to inversion and reflection of the torus cycles, captures the monodromy between primary fields. It is given by 
 \begin{align}
 S_{{\bf a}{\bf b}}=\frac{1}{\mathcal{D}}\sum_{\bf c}d_{\bf c}N_{{\bf a}{\bf b}}^{\bf c}e^{2\pi i(h_{\bf c}-h_{\bf a}-h_{\bf b})} \label{Smatrix}
 \end{align} 
 where $\mathcal{D}=2$ is the total quantum dimension, $d_{\bf a}$ is the quantum dimension for the primary field ${\bf a}$ ( $d_1=d_\gamma=d_{s_\pm}=1$ and $d_\sigma=\sqrt{2}$), and $N_{{\bf a}{\bf b}}^{\bf c}=0,1$ is the number of ways the primary fields ${\bf a}$ and ${\bf b}$ can be fused to ${\bf c}$. In the basis $\{1,\gamma,\sigma\}$, for $N$ odd, and $\{1,\gamma,s_+,s_-\}$, for $N$ even, the $S$ matrices are 
 \begin{align}
 S&=\frac{1}{2}\left(\begin{smallmatrix}1&1&\sqrt{2}\\1&1&-\sqrt{2}\\\sqrt{2}&-\sqrt{2}&0\end{smallmatrix}\right)\quad\mbox{and}\quad S&=\frac{1}{2}\left(\begin{smallmatrix}1&1&1&1\\1&1&-1&-1\\1&-1&i^{N/2}&-i^{N/2}\\1&-1&-i^{N/2}&i^{N/2}\end{smallmatrix}\right),
 \end{align} respectively. The fusion rules ${\bf a}\times{\bf b}=\sum_{\bf c}N_{{\bf a}{\bf b}}^{\bf c}{\bf c}$ can then be read-off from the modular $S$-matrix through the Verlinde formula 
 \begin{align}
 N_{{\bf a}{\bf b}}^{\bf c}=\sum_{\bf d}\frac{\mathcal{S}_{{\bf a}{\bf d}}\mathcal{S}_{{\bf b}{\bf d}}\mathcal{S}^\ast_{{\bf c}{\bf d}}}{\mathcal{S}_{1{\bf d}}}.\label{Verlindeformula}
 \end{align}

From \eqref{spinsigmas}, \eqref{Tmatrix} and \eqref{Smatrix}, the modular information of the $SO(N)_1$ WZW CFT follows a sixteenfold periodicity \begin{align}S_{SO(N)_1}=S_{SO(N+16)_1},\quad T_{SO(N)_1}=T_{SO(N+16)_1}.\end{align} In other words, the set of primary fields, their fusion rules and conformal spins (scaling dimensions) remain the same when $N$ increases by 16. In fact there is a stable equivalence \begin{align}SO(N+16)_1\cong SO(N)_1\otimes(E_8)_1.\label{stbeqper}\end{align} where $E_8$ is the largest exceptional simply-laced Lie algebra. The $E_8$ WZW theory at level 1 has trivial modular content and contains only the trivial primary field. Thus it does not add additional modular content to $SO(N)_1$ in the tensor product \eqref{stbeqper}, but only modifies the central charge by 8 to match between that of $SO(N+16)_1$ and $SO(N)_1$.

%

\bibliographystyle{apsrev}
\bibliography{LoopsLinksBIB}

\end{document}